\documentclass[aps,twocolumn,notitlepage,pra,superscriptaddress,nofootinbib]{revtex4-1}

\usepackage[dvipsnames]{xcolor}
\usepackage{hyperref}
\hypersetup{
  breaklinks=true,
  colorlinks=true,
  allcolors=BlueViolet,
}

\usepackage{amsmath,amssymb,bm}
\usepackage{graphicx}
\usepackage{epstopdf}
\usepackage{latexsym}
\usepackage{enumitem}
\usepackage[normalem]{ulem}
\usepackage{natbib}
\usepackage{braket}
\usepackage{comment}
\usepackage{soul}
\usepackage{listings}

\definecolor{dkgreen}{rgb}{0,0.6,0}
\definecolor{gray}{rgb}{0.5,0.5,0.5}
\definecolor{mauve}{rgb}{0.58,0,0.82}

\newcommand{\densitymat}{\hat{\rho}(\mathbf{x})}

\newcommand{\gammaone}{\gamma^{(1)}}
\newcommand{\gammatwo}{\gamma^{(2)}}

\begin{document}

\title{Variational quantum state preparation for quantum-enhanced metrology in noisy systems}
% \title{Optimal quantum state preparation for metrology with noisy quantum systems}
\author{Juan C. Zu\~{n}iga Castro}
\affiliation{Homer L. Dodge Department of Physics and Astronomy, The University of Oklahoma, Norman, OK 73019, USA}
\affiliation{Center for Quantum Research and Technology, The University of Oklahoma, Norman, OK 73019, USA}
\author{Jeffrey Larson}
\author{Sri Hari Krishna Narayanan}
\affiliation{Argonne National Laboratory, Mathematics and Computer Division, Lemont, IL 60439}
\author{Victor E. Colussi}
\affiliation{Infleqtion, Inc., 3030 Sterling Cir Boulder, CO 80301}
\author{Michael A. Perlin}
\affiliation{Global Technology Applied Research, JPMorgan Chase, New York, NY 10017, USA}
\affiliation{Infleqtion, Inc., 141 West Jackson Blvd Suite 1875 Chicago, IL 60604}
\author{Robert J. Lewis-Swan}
\affiliation{Homer L. Dodge Department of Physics and Astronomy, The University of Oklahoma, Norman, OK 73019, USA}
\affiliation{Center for Quantum Research and Technology, The University of Oklahoma, Norman, OK 73019, USA}
\date{\today}

\begin{abstract}
    We investigate optimized quantum state preparation for quantum metrology applications in noisy environments. Using the QFI-Opt package,
    we simulate a low-depth variational quantum circuit (VQC) composed of a sequence of global rotations and entangling operations applied to a chain of qubits that are subject to dephasing noise. The parameters controlling the VQC are numerically optimized
    %in an effort to prepare optimal quantum states for sensing global spin rotations, using
    to maximize the quantum Fisher information, which characterizes the ultimate metrological sensitivity of a quantum state with respect to a global rotation.
    %, as a cost function.
    We find that regardless of the details of the entangling operation implemented in the VQC, the optimal quantum states can be broadly classified into a trio of qualitative regimes---cat-like, squeezed-like, and product states---
    %separated by the relative dephasing strength.
    associated with different dephasing rates.
    Our findings are relevant for designing optimal state-preparation strategies for next-generation quantum sensors exploiting entanglement, such as time and frequency standards and magnetometers, aimed at achieving state-of-the-art performance in the presence of noise and decoherence. 
\end{abstract}

\maketitle

\section{Introduction}
Quantum sensing holds great promise among emerging technologies that exploit quantum phenomena such as entanglement to achieve a near-term, substantive quantum advantage over state-of-the-art classical devices. Prominent successes include LIGO, which has 
increased its detection volume with frequency-dependent squeezed light
\cite{ligoo4detectorcollaboration2023broadband}, and the use of squeezed quantum states in searches for dark-matter candidates \cite{Haystac_2023}. Similarly, optical frequency metrology experiments are at the precipice of benefiting from quantum entanglement and correlations to push measurement precision beyond the limits of quantum projection noise \cite{Colombo_EntangledClocks_2022,robinson_2024_comparison,bothwell_2022_redshift,Pedrozo_EntanglementOpticalTransition_2020}.

Progress toward the paradigm of quantum-enhanced sensing is being fueled through rapid improvements in technical capabilities, such as the realization of fully controllable intermediate-scale quantum hardware featuring tens to even thousands of high-quality qubits \cite{Ebadi2021_256qubits,Manetsch2024_6100qubits}. These systems exhibit long coherence times, an exquisite degree of coherent single-qubit control, and innovative methods for many-body control (i.e., entangling operations) via natural or engineered interactions \cite{britton2012engineered,Katz_2023_NbodyIons,Bornet_2023_RydbergFloquetSqueezing,Miller_2024_TATpolarmolecules,Pan_2023_SpinExchange}.

A key challenge for current research is to establish robust metrological schemes that exhaust the potential of these hardware platforms in the presence of technical imperfections and decoherence \cite{Jiao2023review,MacLellan2024variational,Campbell2023}. 
In the absence of these, optimal entangled quantum states for metrology can be obtained by analytically maximizing the quantum Fisher information $F_Q$, which bounds from below the achievable variance $(\Delta\theta)^2$ in an estimation parameter $\theta$ via the quantum Cramer--Rao bound, $(\Delta\theta)^2 \geq 1/F_Q$ \cite{caves_1994_qfi,Skotiniotis_2015,Liu_2020_QFIreview}.
% \changed{In the absence of these, optimal entangled quantum states for metrology may be numerically obtained~\cite{toth-vertesi-ppt}, or are often exactly known and can be obtained by analytically maximizing the quantum Fisher information $F_Q$, which bounds from below the achievable variance $(\Delta\theta)^2$ in an estimation parameter $\theta$ via the quantum Cramer-Rao bound, $(\Delta\theta)^2 \geq 1/F_Q$ \cite{caves_1994_qfi,Skotiniotis_2015,Liu_2020_QFIreview}.} 
In the context of sensing spin rotations,  diverse protocols exist, including non-equilibrium and quasi-adiabatic dynamics \cite{lee_2006_cat,Leibfried_2005_cat,zoller_cat_2003,Haine_2018_cat}, to prepare these ideal entangled states as well as optimize signal acquisition and readout to infer $\theta$ \cite{Leibfried_2004_IBR,Haine_2018_IBR,Cao2024-rb,Finkelstein2024-cv,direkci-bayesian-phase-estimation},
%GAIL - I find the "and which" awakward. Does "which" refer to any of the protocols? If so, you might start a new sentence and say Such protocols...
and which can be tailored for specific experimental platforms. However, these optimal quantum states---corresponding to cat-like macroscopic superposition states---are catastrophically susceptible to decoherence \cite{Huelga_1997_decoherence,Pezze_2018_review}, which is ever-present in real-world quantum hardware.
It remains an open challenge to identify and prepare states that maximize quantum Fisher information (QFI) in the presence of decoherence \cite{Brun_2019_mixedqfi}.

In this work we perform a preliminary study of optimal state preparation for metrology in a noisy quantum environment. Our work is motivated by an effort to bring to bear existing resources and coherent control to maximize the potential of near-term quantum devices to demonstrate a state-of-the-art quantum advantage in sensing tasks, as opposed to proof-of-principle realizations. To make our task concrete, we numerically optimize a simple, low-depth variational quantum circuit (VQC; see Fig.~\ref{fig:VQC}) for state preparation in an effort to maximize the resulting quantum Fisher information. Motivated by current physical platforms for quantum computation and simulation, we include a range of entangling gates in our VQC based on a variety of natural or engineered two-qubit interactions. We seek to maximize the QFI of the state generated by the VQC for a range of experimentally relevant noise models. We find that the optimal states can typically be grouped into several regimes---such as cat-like and squeezed-like quantum states---regardless of the nature of the entangling interactions, with some caveats associated with interaction range and system size. While our calculations are limited to relatively small systems (because the QFI for mixed quantum states is computationally expensive to obtain), we comment on implications for larger ensembles of qubits.

Our paper is structured as follows. In Sec.~\ref{sec:vqc} we outline our basic VQC, including fundamental entangling operations and sources of decoherence. We present our results for the optimal QFI obtained with these VQCs in Secs.~\ref{sec:results} and \ref{sec:imbalanced-decoherence}, which includes a detailed analysis of the structure of the obtained optimal quantum states as well as a discussion of the dependence on system size and interaction range. In Sec.~\ref{sec:conclusion} we summarize our findings and present suggestions for future research.

\section{Variational quantum state preparation}\label{sec:vqc}
%VQC outline
\begin{figure}
    \centering
     \includegraphics[width=8.6cm]{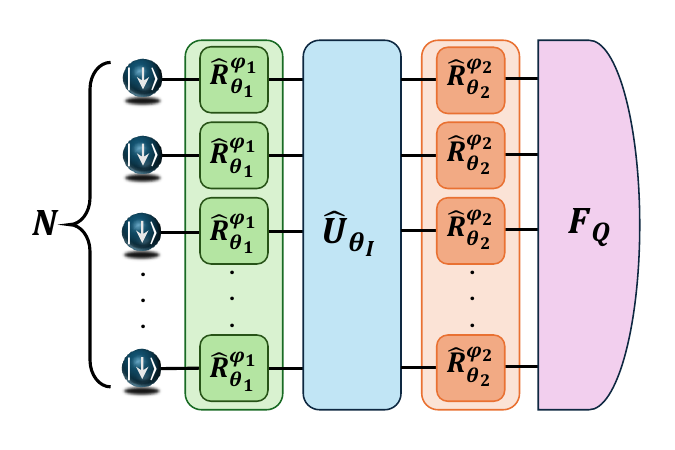}
    \caption{A one-dimensional chain of $N$ qubits (left to right)  prepared in the state $\ket{\downarrow}^{\otimes N}$. The qubits are globally rotated by $\hat{R}_{\theta_1}^{\varphi_1}$ through an angle $\theta_1$ about an axis defined by the azimuthal angle $\varphi_1$. This is followed by application of an entangling gate $\hat{U}_{\theta_{\mathrm{I}}}$ characterized by $\theta_{\mathrm{I}} = \chi t_{\mathrm{I}}$. Then, the qubits are globally rotated by$\hat{R}_{\theta_2}^{\varphi_2}$ through an angle $\theta_2$ about an axis defined by the azimuthal angle $\varphi_2$. We compute the QFI of the state prepared by the VQC, $\densitymat$, which is parameterized by $\mathbf{x} = (\theta_1, \varphi_1, \theta_{\mathrm{I}}, \theta_2, \varphi_2)$. See the main text for a detailed definition of the VQC stages and associated parameters.}
    \label{fig:VQC}
\end{figure}

As a basis for our investigation we consider a low-depth VQC to prepare a quantum state of $N$ qubits to be used for sensing global qubit rotations. The applications for such states are diverse and include frequency metrology and magnetometry. The VQC consists of two key pieces: (i) a set of global rotations occurring during the initial and final stages to prepare and reorient the qubits for sensing and (ii) an intermediate stage where the qubits interact to create entanglement. In more detail (see Fig.~\ref{fig:VQC}), the VQC sequence begins by preparing an ensemble of $N$ qubits in a pure product state $\ket{\psi_0} = \ket{ \downarrow }^{\otimes N}$. The initial and final rotations are parameterized by  $\hat{R}_{\theta_j}^{\varphi_j} = \prod_{k=1}^{N} e^{-i\theta_j\hat{\sigma}^{\varphi_j}_k/2}$, where $\hat{\sigma}_k^{\varphi_j} = \hat{\sigma}^x_k\cos{(\varphi_j)} + \hat{\sigma}_k^y\sin{(\varphi_j)}$ are Pauli matrices for the $k$th qubit in a rotated basis, where $\theta_j$ and $\varphi_j$ respectively define the rotation angle and axis for the initial ($j = 1$) or final ($j=2$) rotation. 
The intermediate entangling step is generated by the gate:
\begin{equation}\label{eqn:general-entangling-gate}
    \hat{U}_{\theta_{\mathrm{I}}} = e^{-it_{\mathrm{I}}\hat{H}_{\mathrm{I}}(\chi)},
\end{equation}
where the Hamiltonian $\hat{H}_{\mathrm{I}}$ describes the interaction of the qubits with characteristic coupling strength $\chi$ (see below). We parameterize the ``strength'' of the entangling gate through the quantity $\theta_{\mathrm{I}} = \chi t_{\mathrm{I}}$, where $t_{\mathrm{I}}$ is the duration of the applied gate. Note that this preliminary description of the entangling gate as a unitary operation ignores the role of decoherence, but we will address this in the following subsections.

In total, the VQC is characterized by a set of five parameters, $\mathbf{x} = (\theta_1, \varphi_1,\theta_I,\theta_2, \varphi_2)$, and additionally through the choice of entangling Hamiltonian, $\hat{H}_{\mathrm{I}}$. We denote the output state of the VQC in terms of these parameters via the density matrix $\hat{\rho}(\mathbf{x})$.

%Spin models
\subsection{Entangling gates}
We study a variety of Hamiltonians $\hat{H}_{\mathrm{I}}$ for the entangling gate, motivated by both conceptual and practical considerations. First, to benchmark our findings, we explore one-axis (OAT) and two-axis (TAT) twisting Hamiltonians featuring infinite-range pairwise interactions between qubits \cite{spin-squeezing-ueda}. The generation of entangled states for metrology using the unitary dynamics generated by these Hamiltonians, such as spin-squeezed and Greenberger--Horne--Zeilinger (GHZ) states, is well established~\cite{spin-squeezing-ueda, qfi-toth,Pezze_2018_review}. In particular, OAT dynamics have been realized and exploited in a diversity of experimental platforms including trapped ions \cite{BohnetJustinG.2016Qsda}, Bose--Einstein condensates \cite{Muessel_2014_OAT}, and cavity-QED \cite{leroux_2010_OAT,hosten_2016_OAT,norcia_2018_OAT,braverman_2019_oat}; and  early work has proposed \cite{he2019engineering} and even demonstrated \cite{Miller_2024_TATpolarmolecules, Luo_2024_cavityTAT} TAT interactions. In addition,  motivated by current efforts to generate entangled states in quantum simulation platforms featuring finite-range pairwise couplings between qubits \cite{perlin2020spin, young2023enhancing, kaufman_2023_squeezing, rey_2023_finiterange, bornet2023scalable}, we also study Ising, XX, and f-TAT (finite-range TAT) spin chain models with power-law interactions.
The Hamiltonians for these models are
\begin{align} \label{eqn:ising-ham} 
    \hat{H}_{\mathrm{Ising}} &= \sum_{i\ne j}{\chi_{ij} \hat{\sigma}_i^z\hat{\sigma}_j^z}, \\ \label{eqn:xx-ham}
    \hat{H}_{\mathrm{XX}} &= \sum_{i\ne j}{\chi_{ij}(\hat{\sigma}_i^x\hat{\sigma}_j^x + \hat{\sigma}_i^y \hat{\sigma}_j^y}), \text{ and}\\ \label{eqn:tat-ham}
    \hat{H}_{\text{f-TAT}} &= \sum_{i\ne j}{\chi_{ij}(\hat{\sigma}_i^x\hat{\sigma}_j^y + \hat{\sigma}_i^y\hat{\sigma}_j^x)}.
\end{align}
Here $i$ and $j$ index individual qubits at positions $r_i$ and $r_j$, and the couplings $\chi_{ij} = \chi/\mathcal{N}_{\alpha}\vert r_i-r_j\vert^{\alpha}$, where $\chi$ sets the strength of interactions and $\mathcal{N}_{\alpha} = (N-1)^{-1}\sum_{i\neq j} \vert r_i - r_j\vert^{-\alpha}$ is a normalization factor to ensure that the Hamiltonian is extensive~\cite{kac,kacDefenuNicolo2021Mads} and simplify comparison of the models as a function of decoherence strength (see below). Throughout this manuscript we set $\hbar = 1$; and, for simplicity, we focus on qubits in a one-dimensional geometry that can be realized using, for example, arrays of neutral atoms or trapped ions.

We will restrict our focus to interactions featuring power-law exponents $\alpha = 0,...,6$, motivated by the range featured in prominent quantum science platforms such as Rydberg atoms ($\alpha = 3,6$), trapped ions ($\alpha \in [0,3]$), molecules ($\alpha = 3$), and light-matter systems ($\alpha = 0$). The limiting case of $\alpha = 0$ corresponds to all-to-all interactions, for which the Ising and XX models reduce (up to a constant of motion) to the OAT Hamiltonian $\propto \hat{S}_z^2$, while f-TAT resolves to the TAT Hamiltonian $\propto(\hat{S}_x\hat{S}_y + \hat{S}_y\hat{S}_x)$.

%Noise models
\subsection{Decoherence}
% Motivated by NISQ hardware...

Motivated by the implementation of the above VQC in real-world hardware, we incorporate into our description decoherence due to, for example, stray background fields and spurious couplings to the environment. In our treatment, however, we will make the simplifying assumption that decoherence occurs only during the application of the entangling gate, with the dynamics of the initial and final rotations taken to be sufficiently rapid that they may be approximated as unitary processes. 

The inclusion of decoherence and dissipation means that
% As we would like to incorporate the effects of dissipation and decoherence during the entangling stage of the VQC, 
the unitary representation of the entangling gate given by Eq.~\eqref{eqn:general-entangling-gate} is inappropriate. Instead, we model this stage of the VQC using a Lindblad master equation, 
\begin{equation}\label{eqn:master-eqn}
\frac{d\hat{\rho}}{dt} = -i[\hat{H}_{\mathrm{I}}, \hat{\rho}] + \sum_{\nu, j} \gamma_{\nu}\left(\hat{L}^{\nu}_j\hat{\rho}\hat{L}^{\nu\dagger}_j - \frac{1}{2}\{\hat{L}^{\nu\dagger}_j\hat{L}^{\nu}_j, \hat{\rho}\}\right),
\end{equation}
where $\hat{H}_\mathrm{I}$ is the chosen entangling Hamiltonian [Eqs.~\eqref{eqn:ising-ham}--\eqref{eqn:tat-ham}] and the jump operators $\hat{L}^{\nu}_{k}$ describe decoherence acting on the $k$th qubit at a rate $\gamma_{\nu}$. In this work we focus on two cases: (i) dephasing with $\hat{L}^{\nu}_{k} = \hat{\sigma}_k^{\nu}/2$ for $\nu \in\set{x, \:y, \:z}$ and (ii) spontaneous decay and dephasing along $\hat{z}$ given by $\hat{L}^{-}_{k} = \hat{\sigma}^{-}_k$ and $\hat{L}^{z}_{k} = \hat{\sigma}_k^{z}/2$ at rates $\gamma_-$ and $\gamma_z$, respectively.
\subsection{Optimization of the quantum Fisher information}
We seek to identify optimal state preparation protocols, within the constraints of the VQC defined above, for metrology of global spin rotations. The quantity capturing the suitability of a generic quantum state $\hat{\rho}$ for this task is the quantum Fisher information \cite{Brun_2014_metrology}: 
\begin{equation}\label{eqn:qfi}
    F_Q[\densitymat ; \hat{G}] = 2\sum_{k,l}\frac{(\lambda_k - \lambda_l)^2}{\lambda_k + \lambda_l}\vert\braket{k|\hat{G}|l}\vert^2. 
\end{equation}
Here $\ket{k}$ and $\lambda_k$ are, respectively, the eigenvectors and corresponding eigenvalues of the density matrix prepared by the VQC; that is,   $\hat{\rho} = \sum_{k} \lambda_k \ket{k}\!\bra{k}$, and $\hat{G}$ is the generator of a unitary transformation $\hat{U}_{\phi} = e^{-i\phi\hat{G}}$ encoding a classical parameter $\phi$. The QFI characterizes the minimum uncertainty with which the classical parameter can be estimated through the quantum Cramer--Rao bound, $(\Delta\phi)^2 \geq 1/F_Q[\densitymat ; \hat{G}]$. Without loss of generality, throughout this work we set $\hat{G} = \hat{S}_z = \sum_j \hat{\sigma}^z_j/2$. 

For an uncorrelated product state of the qubits, the uncertainty is bounded from below by the standard quantum limit (SQL): $(\Delta\phi)^2 \ge 1/N$ (equivalently, $F_Q[\densitymat ; \hat{S}_z] \le N$). This bound can be saturated by a coherent spin state wherein the qubits are collectively polarized along a common axis perpendicular to the rotation about $\hat{z}$, for example, $\ket{\rightarrow}^{\otimes N}$ with $\ket\rightarrow \propto \ket\uparrow + \ket\downarrow$. The introduction of correlations and entanglement can lead to quantum-enhanced sensitivity below the SQL, with a fundamental bound given by the Heisenberg limit (HL): $(\Delta\phi)^2 \ge 1/N^2$ (equivalently, $F_Q[\densitymat ; \hat{S}_z] \le N^2$). It is known that an optimal state for the sensing rotations about $\hat{z}$ that saturates this bound is the macroscopic superposition (``cat'' or GHZ) state~\cite{qfi-toth},
\begin{equation}\label{eqn:ghz-state}
    \ket{\mathrm{GHZ},\Phi} = (\ket{\uparrow}^{\otimes N} + e^{i\Phi} \ket{\downarrow}^{\otimes N})/\sqrt{2}, 
\end{equation}
with arbitrary phase $\Phi$. Strategies to dynamically prepare this state are known and have even been demonstrated for small systems \cite{Leibfried_2005_cat}, although their feasibility for scalable quantum-enhanced sensing is severely limited by the catastrophic sensitivity of the cat state to decoherence \cite{Huelga_1997_decoherence,Foss-FeigM2013Dqco-ising-cat-states}. 

In the following section we instead investigate optimal preparation strategies, using the VQC discussed previously, that take into account the effects of decoherence. For a fixed decoherence strength $\gamma$ we obtain the state $\hat{\rho}(\mathbf{x})$ generated by a given VQC with parameters $\mathbf{x}$.
% \changed{While similar efforts vary the entangling Hamiltonian with the aim to generate specific states~\cite{toth-vertesi-vary-ham}, 
We numerically integrate the master equation \eqref{eqn:master-eqn} for a given choice of entangling Hamiltonian $\hat{H}_{\mathrm{I}}$ and associated global rotations. The optimal preparation strategies defined by parameters $\mathbf{x}_{\mathrm{opt}}$ are obtained by numerically optimizing over $\mathbf{x}$ to maximize the QFI of Eq.~\eqref{eqn:qfi}. Throughout this work we use derivative-free algorithms such as the Nelder--Mead method in SciPy \cite{2020SciPy-NMeth} to obtain $\mathbf{x}_{\mathrm{opt}}$. The QFI-Opt package used to simulate the VQC is an open-source Python package~\cite{qfiopt}.

\section{Optimal state preparation with isotropic dephasing}\label{sec:results}
In this section we present the results of our numerical optimization for the range of entangling Hamiltonians defined in Eqs.~\eqref{eqn:ising-ham}--\eqref{eqn:tat-ham}. Moreover, we focus on $\alpha = 0$, $3$ and $6$ and isotropic dephasing noise such that $\gamma_x = \gamma_y = \gamma_z = \gamma$.
% $(\gamma_x, \: \gamma_y, \: \gamma_z) \equiv (\gamma, \gamma, \gamma)$.
This assumption voids bias in our results due to, for example, robustness of the entangling dynamics for specific representations of the Hamiltonians in Eqs.~\eqref{eqn:ising-ham}--\eqref{eqn:tat-ham}~\cite{britton2012engineered, JurcevicP2014Qeae}. A key result of our investigation is that we observe that the states prepared by the optimal VQC can be typically delineated into three regimes: (i) a \emph{cat-like} regime, characterized by quantum states typified by Eq.~(\ref{eqn:ghz-state}) but more broadly described as a superposition of macroscopically distinct projections, (ii) a \emph{squeezed-like} regime, characterized by states with non-zero polarization (i.e., collective spin projection) and anisotropic projection noise, and (iii) an \emph{uncorrelated} regime, corresponding to a separable coherent spin state of the qubits with sensitivity near the SQL. This classification is based on an analysis of the optimal QFI and abrupt transitions in the associated VQC parameters but is further supported through the examination of auxiliary quantities, including state overlap, distribution functions, and spin squeezing, which provide additional information about the structure of the prepared quantum states.

\begin{figure}
    \centering
    \includegraphics[width=0.9\linewidth]{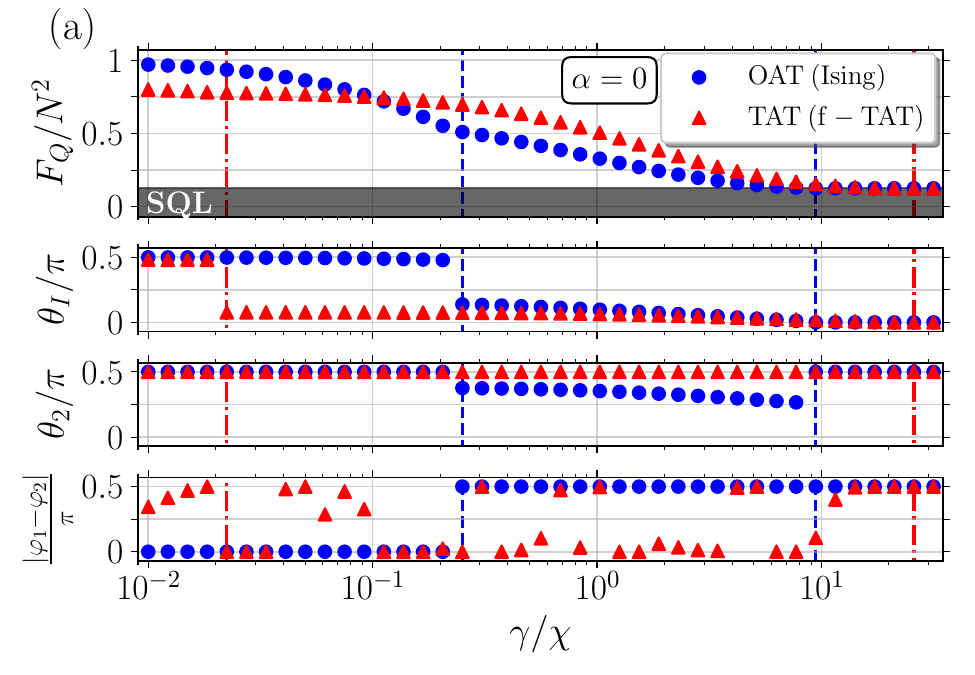}\\
    \includegraphics[width=0.9\linewidth]{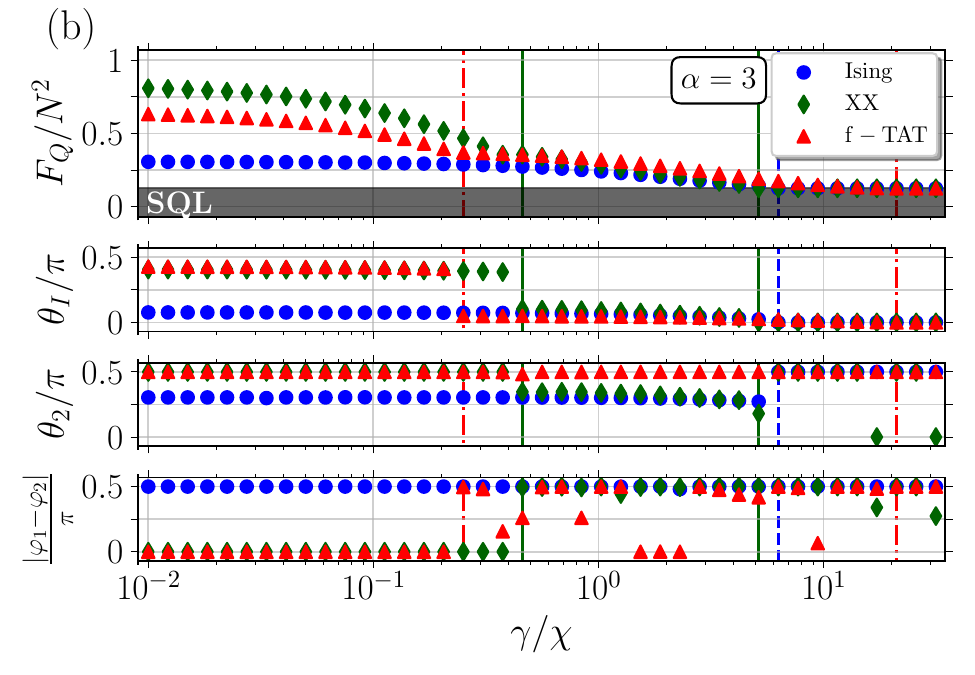}\\
    \includegraphics[width=0.9\linewidth]{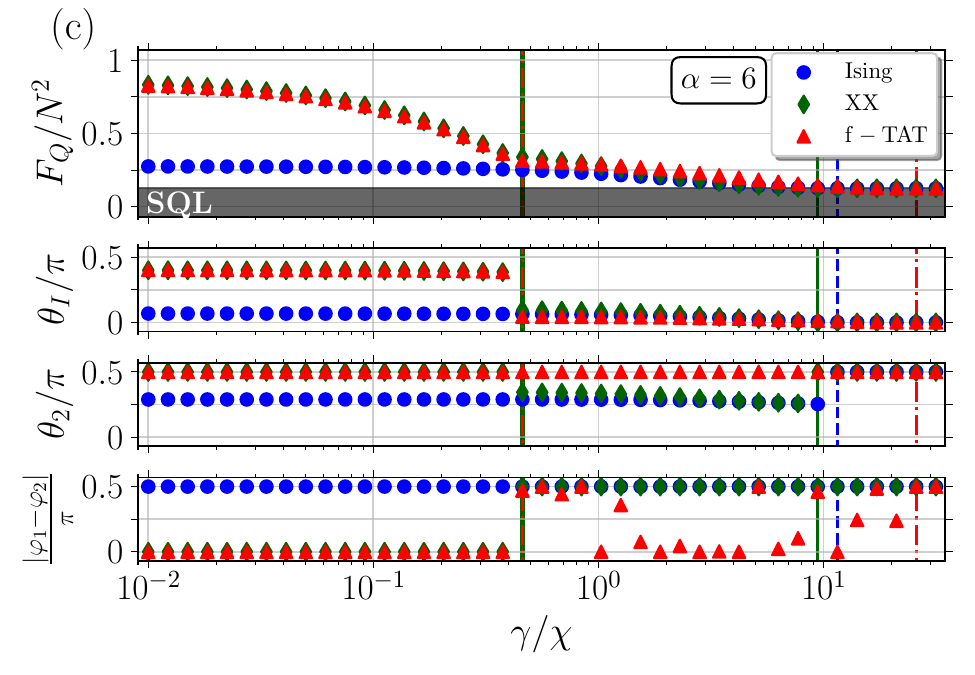}
    \caption{ The normalized QFI $F_Q/N^2$ and optimal VQC parameters $\mathbf{x}_{\mathrm{opt}}$ as functions of isotropic decoherence strength, $\gamma$ (in units of $\chi$) for (a) OAT and TAT and for (b) and (c) Ising, XX, and f-TAT models. In all panels we use $N=8$, and the interaction range is set to (a) $\alpha = 0$, (b) $\alpha = 3$, and (c) $\alpha = 6$. For all cases we exclude the initial rotation angle $\theta_1$, since it is found to identically be either $\pi/2$ (OAT, Ising, and XX) or $0$ (TAT and f-TAT) regardless for nontrivial values of $\gamma < \gamma^{(2)}$ (see main text). 
    The values of $\gamma^{(1)}$ and $\gamma^{(2)}$ for each model are indicated by vertical lines (blue dashed for OAT and Ising, red dot-dashed for TAT and f-TAT, and solid green for XX) to guide the reader between the main regimes. Note that $\gamma^{(1)}$ is absent for the Ising data, while the values of $\gamma^{(1)}$ for XX and f-TAT in panel (c) are indistinguishable.}
    \label{fig:qfi-params-v-gamma}
\end{figure}

\subsection{Infinite-range interactions: One- and two-axis twisting}\label{sec:oat-tat-isotropic}
We begin by presenting results for OAT and TAT with infinite-range interactions ($\alpha =0$), since the dynamics generated by these Hamiltonians is well understood in the absence of noise~\cite{spin-squeezing-ueda, tat-witkowska} and they can provide a template for our later discussion of finite-range interactions. The optimal QFI and parameters obtained for OAT and TAT are plotted in panel (a) of Fig.~\ref{fig:qfi-params-v-gamma} as a function of the isotropic dephasing strength $\gamma$ (in units of $\chi$) and with fixed $N = 8$. For both cases, the QFI begins near the maximal value $F_Q/N^2 \approx 1$ (HL) for $\gamma/\chi \ll 1$ (in fact, it is essentially saturated for OAT) and then monotonically decreases until settling at the SQL ($F_Q/N^2 = 1/N$) for $\gamma/\chi \gg 1$. 

We observe that the QFI appears to be continuous as a function of $\gamma$ for both OAT and TAT, although there is a clear kink in the QFI at $\gamma/\chi\approx 0.25$ for OAT. On the other hand, the associated VQC parameters $\mathbf{x}_{\mathrm{opt}}$ show more abrupt behavior, including discontinuities, that we use to qualitatively classify different regimes of quantum states. In particular, we use the strength of the entangling gate $\theta_{\mathrm{I}}$ to define two characteristic values of $\gamma$. For vanishingly small $\gamma$ we observe that $\theta_I$ takes the largest values ($\theta_{\mathrm{I}}/\pi \approx 0.5$) for both OAT and TAT, but it decreases with increasing $\gamma$. We identify $\gamma^{(1)}$ as the dephasing strength at which $\theta_I$ exhibits a discontinuity and abruptly drops in value, leading to $\gamma^{(1)}/\chi \approx 0.25$ and $0.02$ for OAT and TAT, respectively. Further decrease in $\theta_I$ is then observed for $\gamma > \gamma^{(1)}$ until it vanishes corresponding to turning off the entangling gate. We define the point at which $\theta_{I} = 0$ as $\gamma^{(2)}$ and find $\gamma^{(2)}/\chi \approx 9.44$ and $25.85$ for OAT and TAT, respectively. Additionally, we observe that similar discontinuities or abrupt behavior are present in other VQC parameters, for example, rotation angle $\theta_2$ and the axis difference $\vert \varphi_1 - \varphi_2\vert/\pi$. Note that we do not show data for the initial rotation angle $\theta_1$, since we find that the optimal choice is $\theta_1 = \pi/2$ for OAT and $\theta_1 = 0$ for TAT for $\gamma < \gamma^{(2)}$ (i.e., ``nontrivial'' values of $\gamma$).

Insight into the three distinct parameter regimes carved out by the definition of $\gamma^{(1)}$ and $\gamma^{(2)}$ is found by drilling further into the OAT results, which can be contrasted against the well-understood unitary limit of $\gamma/\chi= 0$. First, when $\gamma < \gamma^{(1)}$, we observe that the optimal parameters $\mathbf{x}_{\mathrm{opt}}$ [see Fig.~\ref{fig:qfi-params-v-gamma} panel (a)] are virtually identical to those that would produce a GHZ state [see Eq.~\eqref{eqn:ghz-state}] in the absence of decoherence \cite{Foss-FeigM2013Dqco-ising-cat-states}. In particular, our optimal VQC sequence corresponds to a first rotation of $\theta_1 = \pi/2$  with an arbitrary value of $\varphi_1$, which orients the initially prepared product state on the collective Bloch sphere equator. The rotation is followed by an entangling gate with strength $\theta_{\mathrm{I}} \approx \pi/2$, which is consistent with the value required to generate a GHZ state using twisting dynamics~\cite{Foss-FeigM2013Dqco-ising-cat-states}. At this point, the generated GHZ-like state is characterized by a collective superposition of qubits oppositely polarized along an axis lying in the equatorial plane (specifically, along an axis perpendicular to that defined by the first rotation about $\varphi_1$). Thus, the final rotation  by an angle $\theta_2 = \pi/2$ about the axis $\varphi_2 \approx \varphi_1$ reorients the GHZ-like state along the $z$-axis [similar to Eq.~\eqref{eqn:ghz-state}]. In the absence of noise, this state features the largest possible fluctuations in the observable $\hat{S}_z$ and thus maximizes the QFI \cite{qfi-toth}.  
As a result, we dub $\gamma < \gamma^{(1)}$ as the \emph{cat-like} regime, since our results suggest that decoherence is sufficiently weak  that the optimal state preparation strategy is simply to closely (i.e., cat-likely) follow the approach established in the absence of decoherence.

On the other hand, in the regime where $\gamma^{(1)} \leq \gamma < \gamma^{(2)}$, $\theta_{\mathrm{I}}$ takes on relatively smaller values that can be associated with the paradigmatic generation of spin-squeezed states~\cite{spin-squeezing-ueda}. Particular to one-axis twisting, the unitary generation of spin-squeezed states occurs via ``shearing'' of the quantum fluctuations of an initial coherent spin state polarized along the equator of the collective Bloch sphere. The resulting squeezing of the projection noise is thus in a spin quadrature that is rotationally offset from the equatorial plane of the Bloch sphere, by an amount that depends on the strength/duration of the twisting interaction ($\theta_I$), and thus requires a final rotation about an axis aligned along the collective polarization of the state. We see signatures of these features in our dynamics with decoherence, including that the final rotation axis is perpendicular to the initial rotation (and thus parallel to the polarization of the state post first rotation), $\vert \varphi_2 - \varphi_1\vert \approx \pi/2$, while the angle of the rotation $\theta_2$ varies with the duration of the entangling interaction characterized by $\theta_{\mathrm{I}}$~\cite{spin-squeezing-ueda}. Thus, since our results suggest that the states within this regime may be closely associated with spin squeezing, we designate $\gamma^{(1)} \leq \gamma < \gamma^{(2)}$ as the \emph{squeezed-like} regime.

\begin{figure}
    \centering
    \includegraphics[width=0.9\linewidth]{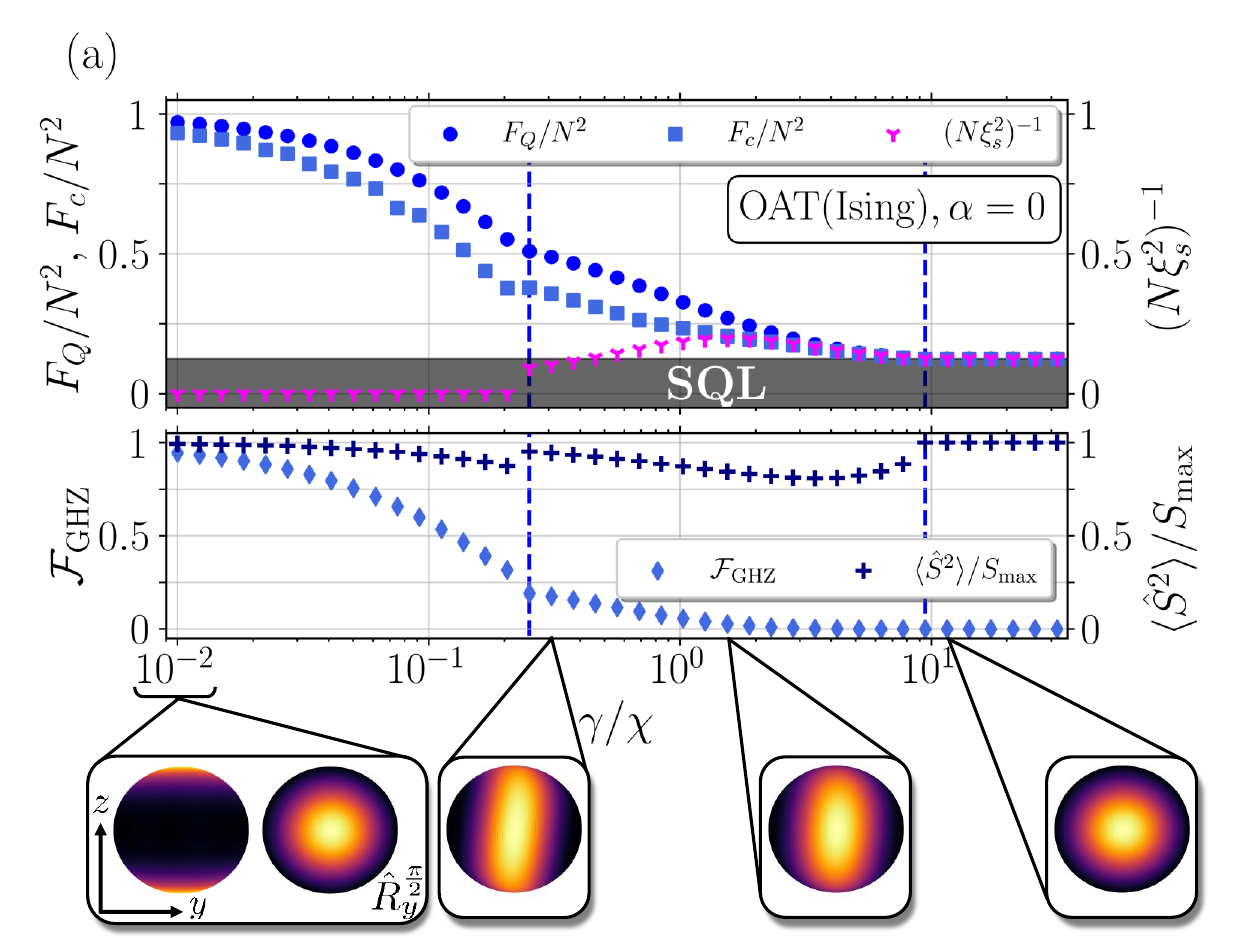}\\
    \includegraphics[width=0.9\linewidth]{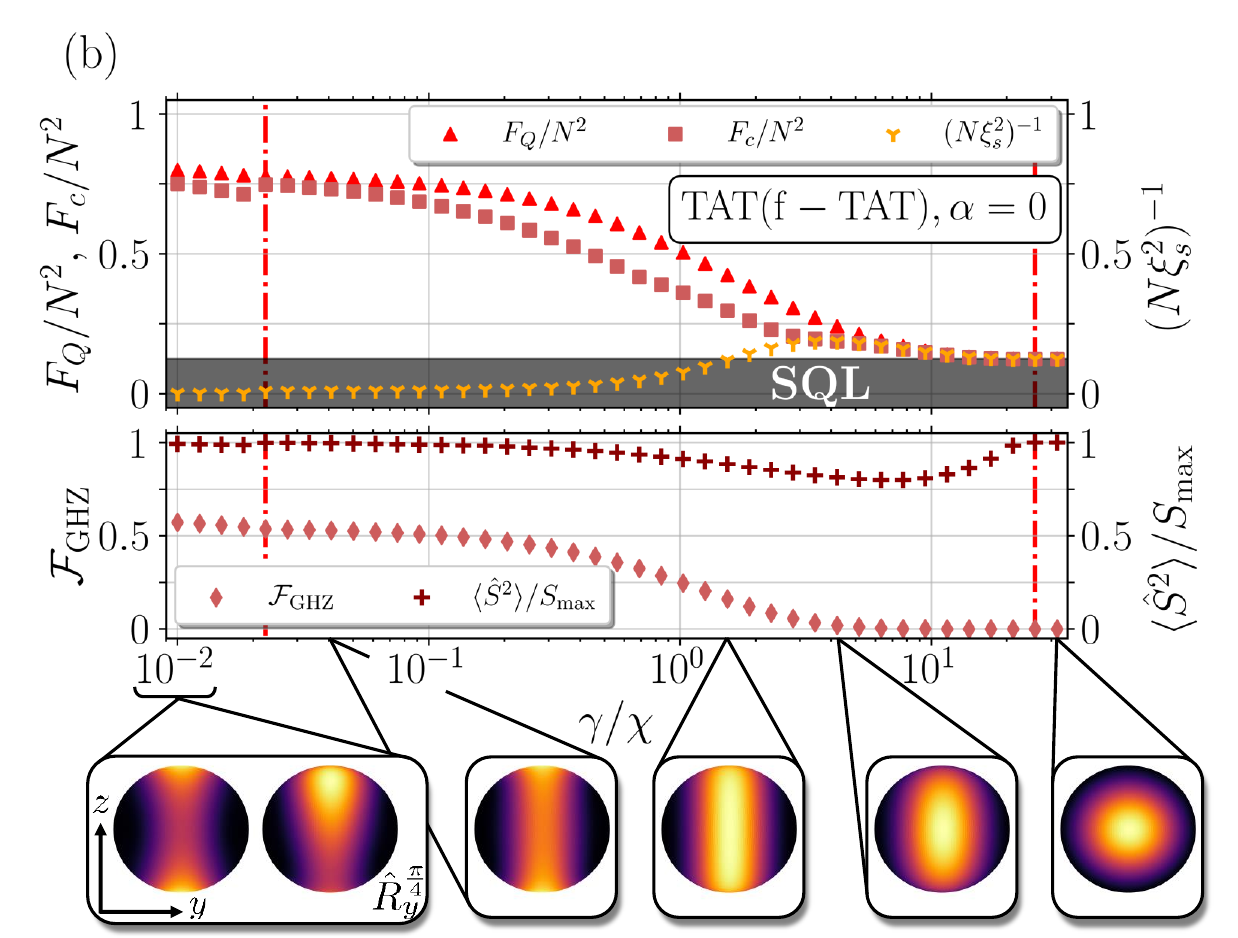}
    \caption{ 
    Diagnostic quantities calculated for optimally prepared states as functions of isotropic dephasing strength $\gamma$ (in units of $\chi$) for (a) OAT and (b) TAT (i.e., Ising and f-TAT Hamiltonians, respectively, with $\alpha = 0$). All data is for $N = 8$. Top subplots: Normalized classical Fisher information ($F_c/N^2$, squares), inverse Wineland squeezing parameter $(N\xi^2_s)^{-1}$, tri-prongs) and the normalized QFI ($F_Q / N^2$, circles for OAT and triangles for TAT). Lower subplots: Overlap of the optimal state $\hat{\rho}(\mathbf{x}_{\mathrm{opt}})$ with a GHZ state ($\mathcal{F}_{\mathrm{GHZ}}$, diamonds) and the normalized collectivity ($\langle \hat{S}^2 \rangle/S^2_{\mathrm{max}}$, crosses). 
    Vertical lines in all panels indicate $\gamma^{(1)}$ and $\gamma^{(2)}$. Below each set of panels we show Husimi phase space distributions for representative quantum states in each regime (see Appendix \ref{sec:husimi} for details of distribution). Bright regions indicate larger probability density. For $\gamma/\chi= 0.01$ we show two perspectives of the same quantum state, rotated for indicated axes and angles.}
    \label{fig:rel-quantities-v-gamma}
\end{figure}

Finally, we denote $\gamma \geq \gamma^{(2)}$ as the \emph{uncorrelated} regime. In this case, since the entangling gate is never applied ($\theta_I = 0$), the VQC reduces to a pair of global rotations that together simply reorient the initial pure, uncorrelated coherent spin state onto some axis in the $x$-$y$ plane to maximize the QFI at the SQL, $F_Q/N^2 = 1/N$. This regime arises for overwhelmingly large decoherence, such that any application of the entangling gate would in fact lead to a QFI below the SQL due to an increase in total projection noise. In other words,  the interactions cannot generate metrologically useful correlations faster than decoherence degrades the quantum state [see the bottom subplot of Fig.~\ref{fig:rel-quantities-v-gamma}].

The behavior of the optimal parameters $\mathbf{x}_{\mathrm{opt}}$ for TAT (also shown in Fig.~\ref{fig:qfi-params-v-gamma}) follows similar trends and is suggestive of a similar classification into three state preparation regimes. In this case, however, the increased complexity of the unitary dynamics for TAT \cite{tat-witkowska} makes quantitative analysis of the parameters more difficult. Thus, in the following we build on the intuition developed from OAT and perform a more systematic investigation of the structure of the prepared quantum states using a variety of diagnostic quantities.

\subsubsection{Characterization of cat-like states}
We first focus on the cat-like regime defined by $\gamma < \gamma^{(1)}$. Motivated by the observation that the QFI in this regime is near the HL ($F_Q/N^2 = 1$) and informed by the close proximity of the parameters $\mathbf{x}_{\mathrm{opt}}$ for OAT to those that are known to generate GHZ states [see Eq.~\eqref{eqn:ghz-state}] for $\gamma/\chi= 0$, we calculate the overlap between the optimally prepared state $\hat{\rho}(\mathbf{x}_\mathrm{opt})$ and a GHZ state. Specifically, we compute the maximum fidelity,
% \begin{equation}
%     \mathcal{F}_{\mathrm{GHZ}} = \left(\mathrm{Tr}\left[\sqrt{\sqrt{\hat{\rho}_{\mathrm{GHZ}}^{\phi}} \:\hat{\rho}(\mathbf{x}_{\mathrm{opt}}) \sqrt{\hat{\rho}_{\mathrm{GHZ}}^{\phi}}}\:\right]\right)^2,
% \end{equation}
\begin{equation}
    \mathcal{F}_{\mathrm{GHZ}} = \max_{\Phi} ~\vert\braket{\mathrm{GHZ},\Phi | \hat{\rho}(\mathbf{x}_{\mathrm{opt}}) | \mathrm{GHZ},\Phi}\vert^2.
\end{equation}
The results for this quantity as a function of $\gamma/\chi$ are shown in the middle panels of Fig.~\ref{fig:rel-quantities-v-gamma}(a) and (b). We observe appreciable fidelities $1/2 < \mathcal{F}_{\mathrm{GHZ}} \leq 1$ in the regime $\gamma < \gamma^{(1)}$ for both OAT [panel (a)] and TAT [panel (b)]. The values for the fidelity are indicative of strong overlap with the GHZ state and the presence of $N$-qubit entanglement \cite{Leibfried_2005_cat}. 
% For both OAT and TAT the fidelity remains appreciable when $\gamma > \gamma^{(1)}$, although it steadily decreases \cite{Huelga_1997_decoherence}. 
Relative to the OAT results, the TAT case is observed to typically feature a reduced fidelity to the GHZ state, indicating that an ideal GHZ is never formed even for $\gamma/\chi\to 0$. Interesting, however, $\mathcal{F}_{\mathrm{GHZ}}$ remains robustly above or quite near $1/2$ even for some values of $\gamma > \gamma^{(1)}$, although it steadily decreases. We attribute this to the inherent richness of states that can generated by TAT interactions \cite{tat-witkowska} and the relatively faster generation of entanglement that is expected \cite{spin-squeezing-ueda}. In contrast, the fidelity for OAT rapidly falls below $1/2$ prior to $\gamma = \gamma^{(1)}$. However, we qualify these observations by noting that the system size used in our calculations is quite small. 

The fidelity is qualitatively supported by visualization of the quantum states in this regime using an averaged Husimi probability distribution (see Appendix \ref{sec:husimi}), shown below the main panels of Fig.~\ref{fig:qfi-params-v-gamma}. These distributions clearly demonstrate that the quantum state is composed of a superposition of two components localized at the north and south pole of the Bloch sphere, consistent with a GHZ-like state.

\subsubsection{Characterization of squeezed-like states}\label{sec:sq-like-characterization}
In the squeezed-like regime found for larger decoherence, $\gamma^{(1)} \leq \gamma < \gamma^{(2)}$, we expect the prepared states to have a non-zero collective polarization and anisotropic quantum fluctuations in the spin quadratures defined perpendicular to the orientation of the state. Mathematically, this structure can be captured by the Wineland squeezing parameter~\cite{wineland-spin-squeezing},
\begin{equation}\label{eqn:wineland-squeezing-witness}
    \xi^2_s = N \min_{\mathbf{n}_{\perp}} \frac{\langle (\Delta \hat{S}_{\mathbf{n}_{\perp}})^2 \rangle}{\vert \langle \hat{\mathbf{S}} \rangle \vert^2},
\end{equation} 
where $\langle \hat{\mathbf{S}} \rangle = (\langle \hat{S}_x \rangle , \langle \: \hat{S}_y \rangle,\:  \langle\hat{S}_z\rangle)$ is the collective Bloch vector of the state and $\mathbf{n}_{\perp}$ denotes a unit vector lying in a plane orthogonal to $\langle \hat{\mathbf{S}} \rangle$. The squeezing parameter is obtained by minimizing the fluctuations $\langle (\Delta \hat{S}_{\mathbf{n}_{\perp}})^2 \rangle$ with respect to the orientation of $\mathbf{n}_{\perp}$. A value $\xi^2_s < 1$ reflects that the quantum fluctuations along one spin quadrature are squeezed below the projection noise limit such that the state can enable sensing of spin rotations below the SQL using a simple Ramsey sequence \cite{wineland-spin-squeezing}. More generally, squeezing sets a lower bound on the QFI (and thus the metrological utility of the quantum state) through the inequality $F_Q/N^2 \geq (N\xi^2_s)^{-1}$.

We plot the inverse squeezing parameter $(N\xi^2_s)^{-1}$ as a function of $\gamma$ in the upper panels of Fig.~\ref{fig:rel-quantities-v-gamma}(a) and (b). We observe that $(N\xi^2_s)^{-1} = 0$ in the cat-like regime of $\gamma < \gamma^{(1)}$ as the polarization of the cat state vanishes; that is, $\langle \hat{\mathbf{S}} \rangle = 0$. On the other hand, for $\gamma \geq \gamma^{(1)}$ we observe $(N\xi^2_s)^{-1} > 0$. Initially, as $\gamma$ is increased past this transition, the inverse squeezing parameter does not breach the SQL, that is, $(N\xi^2_s)^{-1} < 1/N$, indicating that while the QFI is large ($F_Q/N^2 > 1/N$) and corresponds to a quantum-enhanced sensitivity, the state does not exhibit simple squeezing that can be exploited using Ramsey interferometry with mean values of global spin measurements. We attribute this to \emph{oversqueezing}, which has been studied in the context of OAT dynamics \cite{strobel_2014_qfi,BohnetJustinG.2016Qsda}. To demonstrate this, we show example Husimi distributions for representative quantum states in this regime below panels in Fig.~\ref{fig:rel-quantities-v-gamma}(a) and (b). These clearly illustrate that the quantum state features anisotropic quantum noise. However, the distribution wraps appreciably around the Bloch sphere, leading to a shrinking of the collective Bloch vector $\langle \hat{\mathbf{S}} \rangle$ and the failure of the squeezing parameter to capture the metrological sensitivity of the state. As $\gamma$ further increases, the squeezing parameter breaks through the SQL and eventually matches the QFI, $(N\xi^2_s)^{-1} = F_Q/N^2$. In this latter regime, the metrological utility of the state is thus entirely captured by spin squeezing.

\subsubsection{Additional diagnostic quantities}
In addition to the fidelity to the GHZ state and the squeezing parameter, it is insightful to consider other quantities. First, we compute the \emph{collectivity} of the prepared states (see Fig.~\ref{fig:rel-quantities-v-gamma}), $\langle \hat{S}^2\rangle$, which should be compared against the maximal value of $S^2_{\mathrm{max}} = S(S+1)$ with $S = N/2$ for a collective state. Both OAT and TAT unitary dynamics preserve the collectivity since they feature infinite-range interactions, so that any deviation from the maximum value of $\langle \hat{S}^2\rangle$ should be attributable to the single-qubit dephasing noise. The collectivity is most useful when examined alongside a second quantity, the \emph{classical} Fisher information (CFI), which can be obtained as
\begin{multline}
    % F_c(\theta) = \\4\lim_{\theta\to0}{{\sum_{m_z=-N/2}^{N/2}{\frac{\left(\sqrt{P(m_z;\theta)} - \sqrt{P(m_z;\theta=0)}\right)^2}{{\theta^2}}}}}
    F_c(\theta) = \lim_{\theta\to0} \frac{4}{\theta^2} \sum_{m_z} \left(\sqrt{P(m_z;\theta)} - \sqrt{P(m_z;\theta=0)}\right)^2,
\end{multline}
where $P(m_z)$ is the distribution function for the collective spin projection $m_z$ along $\hat{z}$ and $F_c$ is bounded from above by the QFI, $F_c \leq F_Q$. In the case of unitary OAT and TAT dynamics, the CFI always saturates the QFI when optimized over all possible collective measurements. However, we observe that $F_c$ is less than the QFI across parts of both the cat-like and squeezed-like regimes, particularly when the collectivity strays appreciably from $S^2_{\mathrm{max}}$. This suggests that more sophisticated measurements, such as local (site-resolved) spin observables, are necessary to exhaust the metrological potential of the states we preserve and, in particular, highlights that states in the ``squeezed'' regime---a name that is motivated by a simple analysis of \emph{collective} spin fluctuations---require more nuanced treatment in general.

\subsection{Finite-range interactions: Ising, XX, f-TAT}

\begin{figure*}
    \centering
    \includegraphics[width=0.45\linewidth]{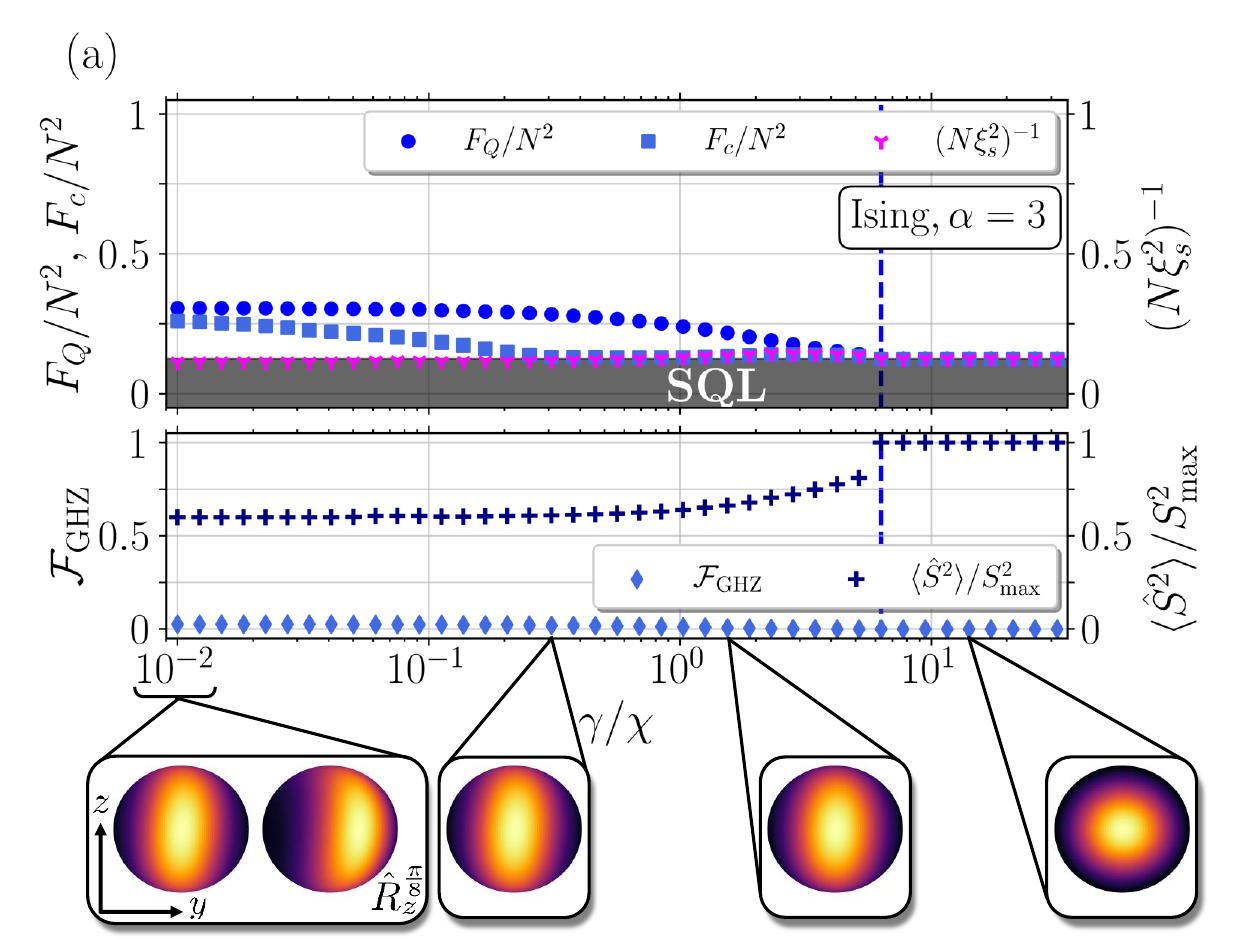}\:\:\:\:\:\:
    \includegraphics[width=0.45\linewidth]{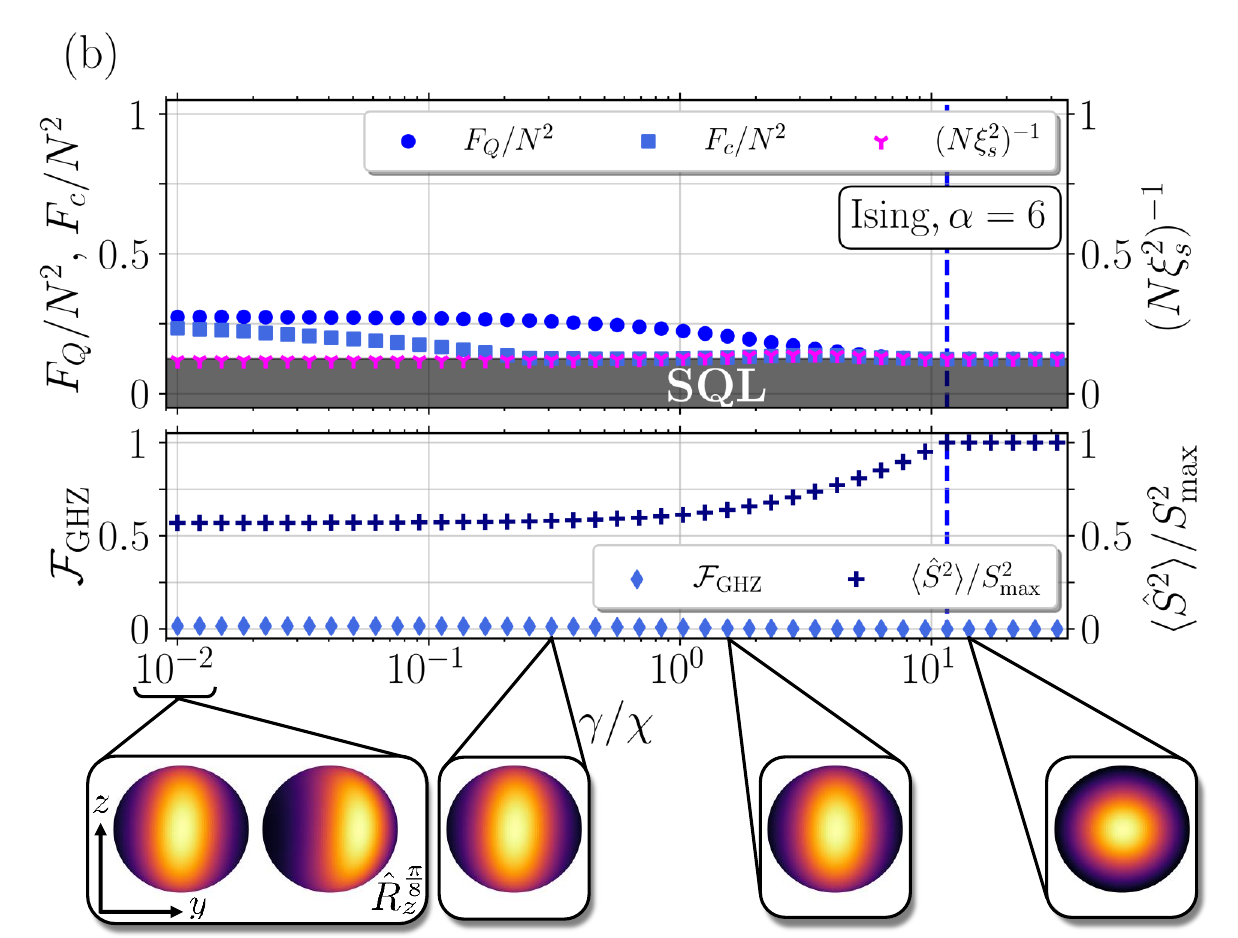}\\
    \includegraphics[width=0.45\linewidth]{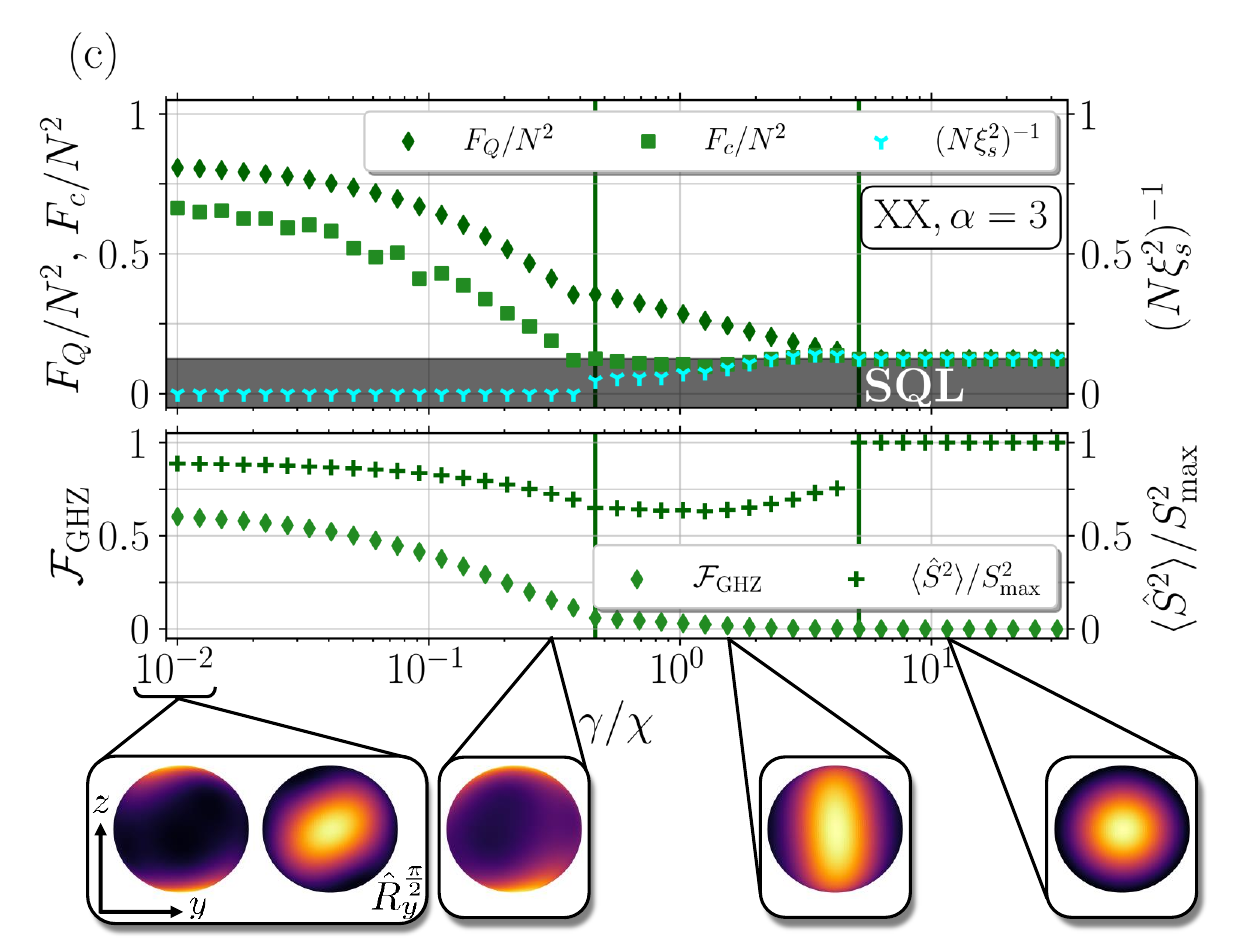}\:\:\:\:\:\:
    \includegraphics[width=0.45\linewidth]{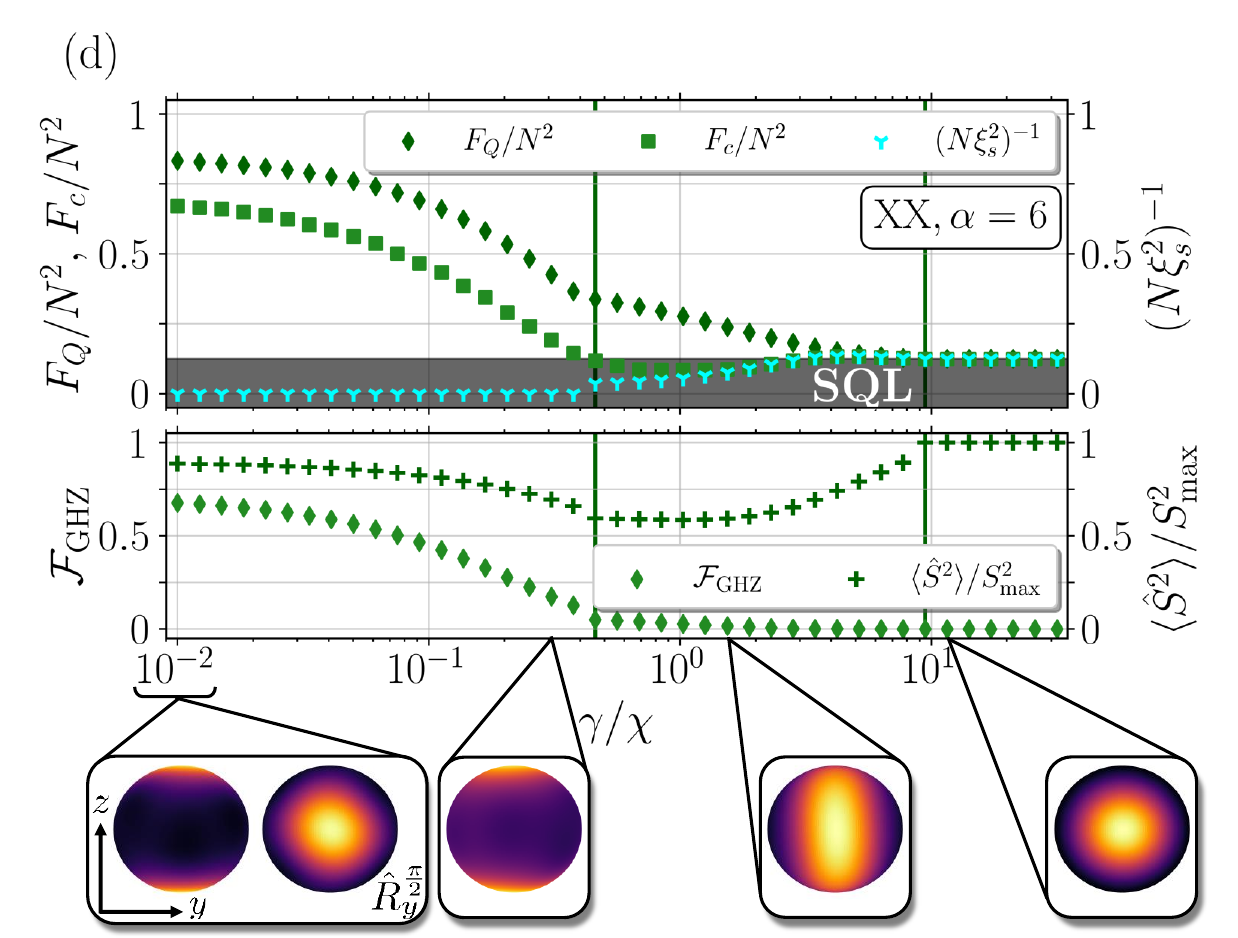}\\
    \includegraphics[width=0.45\linewidth]{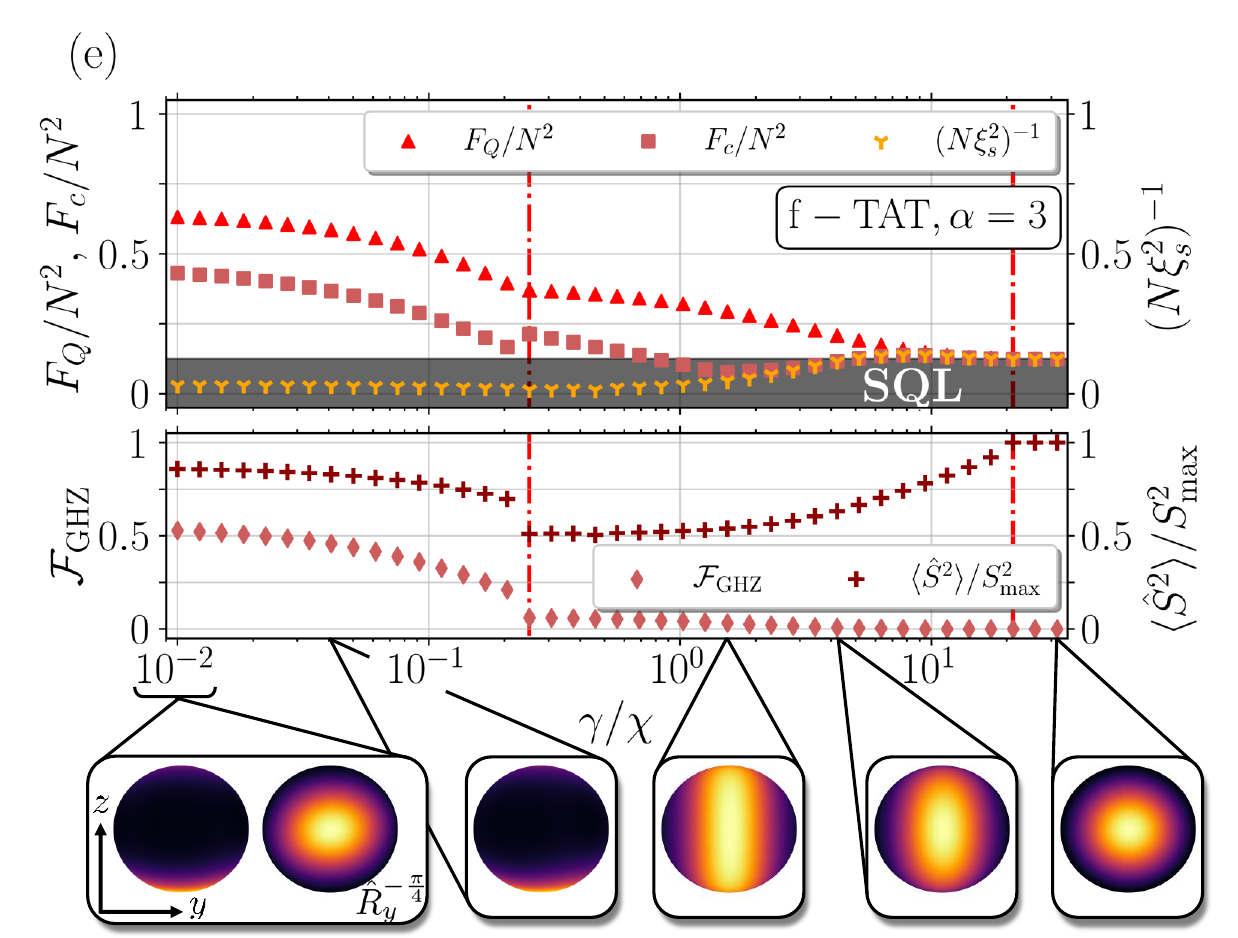}\:\:\:\:\:\:
    \includegraphics[width=0.45\linewidth]{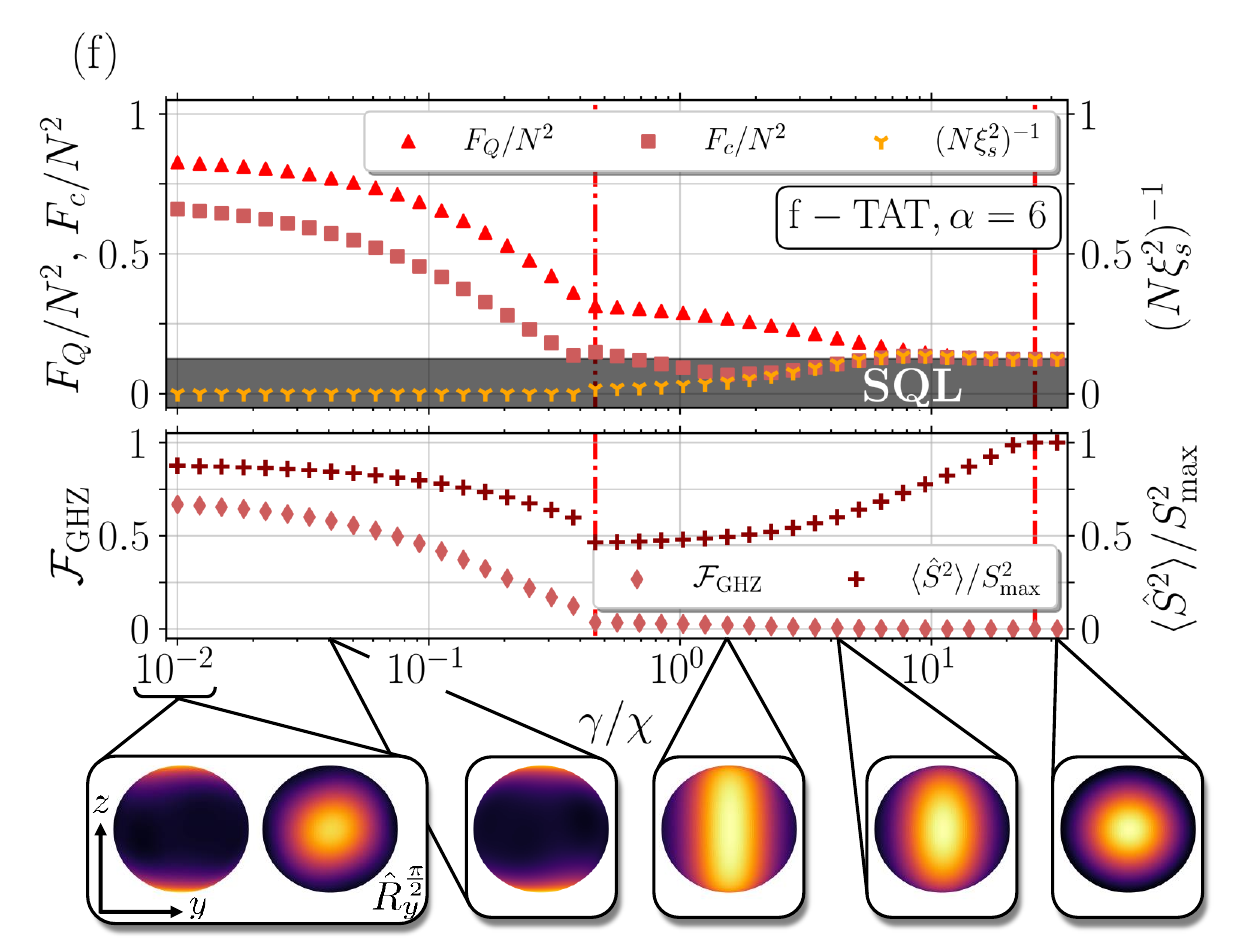}

    \caption{Diagnostic quantities calculated for optimally prepared states as functions of relative dephasing strength $\gamma/\chi$ for (a)--(b) Ising, (c)--(d) XX, and (e)--(f) f-TAT Hamiltonians. All data is calculated for $N = 8$ qubits. Panels (a), (c), and (e) are generated with $\alpha = 3$ whereas panels (b), (d), and (f) use $\alpha = 6$. Top subplots: Normalized classical Fisher information ($F_c/N^2$, squares), inverse Wineland squeezing parameter $(N\xi^2_s)^{-1}$, tri-prongs), and the normalized QFI ($F_Q / N^2$, circles, diamonds ,and triangles, respectively). Lower subplots: Overlap of the optimal state $\hat{\rho}(\mathbf{x}_{\mathrm{opt}})$ with a GHZ state ($\mathcal{F}_{\mathrm{GHZ}}$, diamonds) and the normalized collectivity ($\langle \hat{S}^2 \rangle/S^2_{\mathrm{max}}$, crosses). Vertical lines in all panels indicate $\gamma^{(1)}/\chi$ [absent for (a) and (b)] and $\gamma^{(2)}/\chi$.
    Below each set of panels we show Husimi phase space distributions for representative quantum states in each regime (see Appendix \ref{sec:husimi} for details of distribution). Bright regions indicate larger probability density. For $\gamma/\chi= 0.01$ we show two perspectives of the same quantum state, rotated for indicated axes and angles. } 
    % (a)-(c) down the first column and (d)-(c) down the second column for simple reference in text.}}
    \label{fig:finite-range-rel-quantities-v-gamma}
\end{figure*}

Equipped with an understanding of the results for the infinite-ranged OAT and TAT entangling Hamiltonians, we now generalize to the finite-ranged models given in Eqs.~\eqref{eqn:ising-ham}--\eqref{eqn:tat-ham}. For simplicity, wefocus on the specific coupling exponents $\alpha = 3$ and $6$ in the following. 
% and will present additional results investigating the role of interaction range in a more detailed follow-up work \cite{JZC_2024_upcoming}. 

In Figs.~\ref{fig:qfi-params-v-gamma}(b) and (c), we plot the maximal QFI and optimal parameters $\mathbf{x}_{\mathrm{opt}}$ for these coupling exponents as a function of dephasing strength $\gamma$ with $N = 8$ qubits. Overall, we observe a similar delineation of the QFI and $\mathbf{x}_{\mathrm{opt}}$ into different regimes set by the dephasing strength, with some important caveats. First, for the XX and f-TAT models, we can define an analogous cat-like regime for $\gamma < \gamma^{(1)}$ (see vertical lines in panels), where the transition $\gamma^{(1)}/\chi$ is again defined by an abrupt drop in the interaction time characterized by $\theta_{I}$ and similarly sharp changes in the rotation angles/axes. Similarly, we define a squeezed-like regime for $\gamma^{(1)} \leq \gamma < \gamma^{(2)}$, with the upper bound on the dephasing strength $\gamma^{(2)}/\chi$ set via the uncorrelated regime that emerges when $\theta_I$ finally reaches zero. On the other hand, the Ising model presents distinct behavior. In particular, we observe an absence of the cat-like regime, as the VQC parameters change smoothly throughout a large squeezed-like regime $\gamma < \gamma^{(2)}$, until $\theta_I$ vanishes at the boundary of the uncorrelated regime that exists for $\gamma \geq \gamma^{(2)}$. We note that the extent of the cat-like (where applicable) and squeezed-like regimes is slightly reduced with $\alpha = 3$ when compared with $\alpha = 6$.

While our preliminary identification of the cat-like, squeezed-like, and uncorrelated regimes is supported by an analysis of the optimal parameters $\mathbf{x}_{\mathrm{opt}}$, which show some qualitative similarities to OAT (Ising and XX) and TAT (f-TAT),  we provide stronger evidence through quantities such as the overlap with a GHZ state and the squeezing parameter, which are plotted as a function of $\gamma/\chi$ in Fig.~\ref{fig:finite-range-rel-quantities-v-gamma}.
% \Com{Figs.}~\ref{fig:finite-range-relevant-quantities-alpha3-v-gamma} and \ref{fig:finite-range-relevant-quantities-alpha6-v-gamma}. 
% \changed{Across these figures,  we observe a slight reduction in the extent of the cat-like and squeezed-like regimes for Ising, XX, and f-TAT with $\alpha = 3$ when compared to $\alpha = 6$.} 

The cat-like regime for the XX and f-TAT models 
% [see panels (b) and (c) of \Com{Figs.}~\ref{fig:finite-range-relevant-quantities-alpha3-v-gamma} and~\ref{fig:finite-range-relevant-quantities-alpha6-v-gamma}] 
[see panels (c)-(f) of Fig.~\ref{fig:finite-range-rel-quantities-v-gamma}] is supported by an appreciable overlap with the GHZ state \cite{perlin2020spin}. In particular, for $\gamma/\chi\to 0$ both XX and f-TAT feature $\mathcal{F}_{\mathrm{GHZ}} > 1/2$, indicating the generation of a quasi-collective state that shows signatures consistent with a GHZ-like state. Unlike the infinite-range TAT results, however, the overlap for f-TAT dips below the value of $1/2$ before the transition to the squeezed-like regime. We also observe that the maximum value of $\mathcal{F}_{\mathrm{GHZ}}$ and indeed the QFI $F_Q/N^2$ attained for the XX and f-TAT Hamiltonians are surprisingly greater in the case $\alpha = 6$ rather than the longer-range case with $\alpha = 3$. This distinction is discussed in further detail momentarily. 
In addition to the GHZ fidelity, we highlight that the characteristic collectivity $\langle \hat{S}^2 \rangle$ of the states generated in the cat-like regime remains a substantial fraction of the maximum value $S^2_{\mathrm{max}}$. In contrast, the Ising model [see panels (a) and (b) of Fig.~\ref{fig:finite-range-rel-quantities-v-gamma}] never features a meaningfully large overlap with the GHZ state, even for $\gamma/\chi\to 0$, although we note that the value of the collectivity $\langle \hat{S}^2 \rangle$ remains appreciable, much like the states prepared by XX and f-TAT. Furthermore, unlike the XX and f-TAT examples, for the Ising Hamiltonian the attainable QFI in this regime is marginally reduced for $\alpha = 6$ relative to that of $\alpha = 3$. 

%Squeezed-like...
The squeezed-like regime between $\gammaone \leq \gamma < \gammatwo$ is also significantly more complex for the finite-ranged models. While the inverse squeezing parameter $(N\xi^2_s)^{-1}$ continues to take on non-zero values for $\gamma \geq \gammaone$, reflecting the transition from the GHZ state to one with non-zero collective polarization of the qubits, in contrast to the results for the infinite-ranged models we observe that the squeezing parameter only marginally breaches the SQL (i.e., $(N\xi^2_s)^{-1} < 1/N$) for limited values of $\gamma$. This reflects that while these states may exhibit anisotropic fluctuations/correlations analogous to a spin-squeezed state (see averaged Husimi distributions shown beneath each panel in Fig.~\ref{fig:finite-range-rel-quantities-v-gamma}), the finite-range interactions lead to states that are far from collective. Notably, the collectivity parameter $\langle \hat{S}^2 \rangle$ is far smaller in the squeezed-like regime for these models than previous observations for the infinite-ranged models in Fig.~\ref{fig:rel-quantities-v-gamma}. 

The complexity of the squeezed-like regime is further supported by computing the CFI. For the Ising model, the CFI lies above the SQL for $\gamma \ll \gamma^{(2)}$  but quickly deviates below the QFI as $\gamma$ increases. On the other hand, for the XX and f-TAT models the CFI  shows only a handful of points in the squeezed-like regime where the SQL is breached and does not approach the QFI until the point at which quantum enhancement is essentially lost.
Given that the CFI  we compute is constrained to collective observables, this suggests that, in general, harnessing the QFI of states generated by finite-ranged interactions will require measurement of site-resolved observables or correlations.

% The squeezed-like regime is indicated by values of $(N\xi^2_s)^{-1}>0$. In XX and f-TAT, we observe values of $(N\xi^2_s)^{-1} < 1/N$ indicative of an associated oversqueezed regime [see discussion of Sec.~\ref{sec:sq-like-characterization}]. Although this is predominantly true for Ising, the difference between $(N\xi^2_s)^{-1}$ and the SQL are insignificantly marginal. For larger values of $\gamma/\chi$, a small window with $(N\xi^2_s)^{-1} > 1/N$ (closely associated with an increase in collectivity and $F_c$ and a decrease in overlap with a GHZ) emerges before settling at the SQL. Finally the uncorrelated regime is discerned by QFIs and $(N\xi^2_s)^{-1}$ saturating the SQL and collectivities at maximal values.

\begin{figure}
    \centering
    \includegraphics[width=0.8\linewidth]{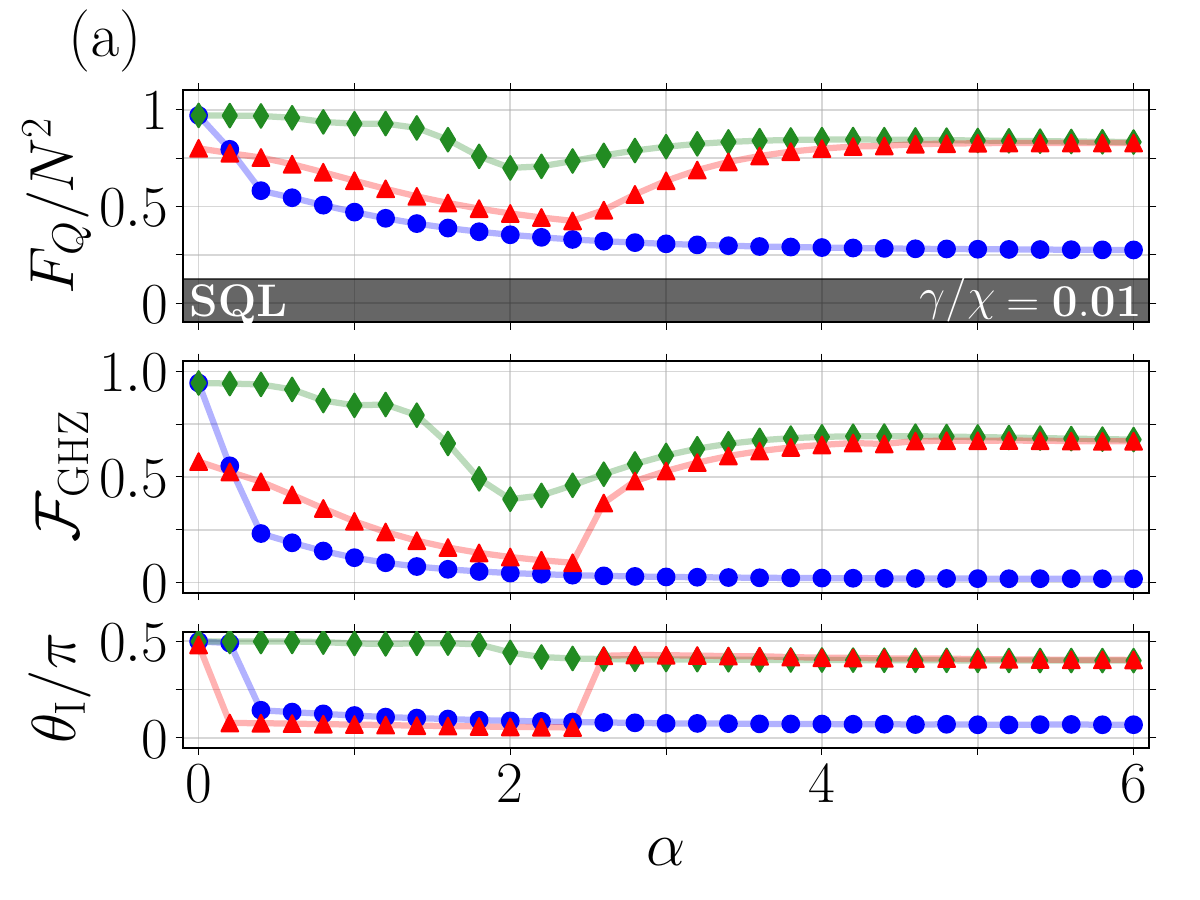}\\
    \includegraphics[width=0.8\linewidth]{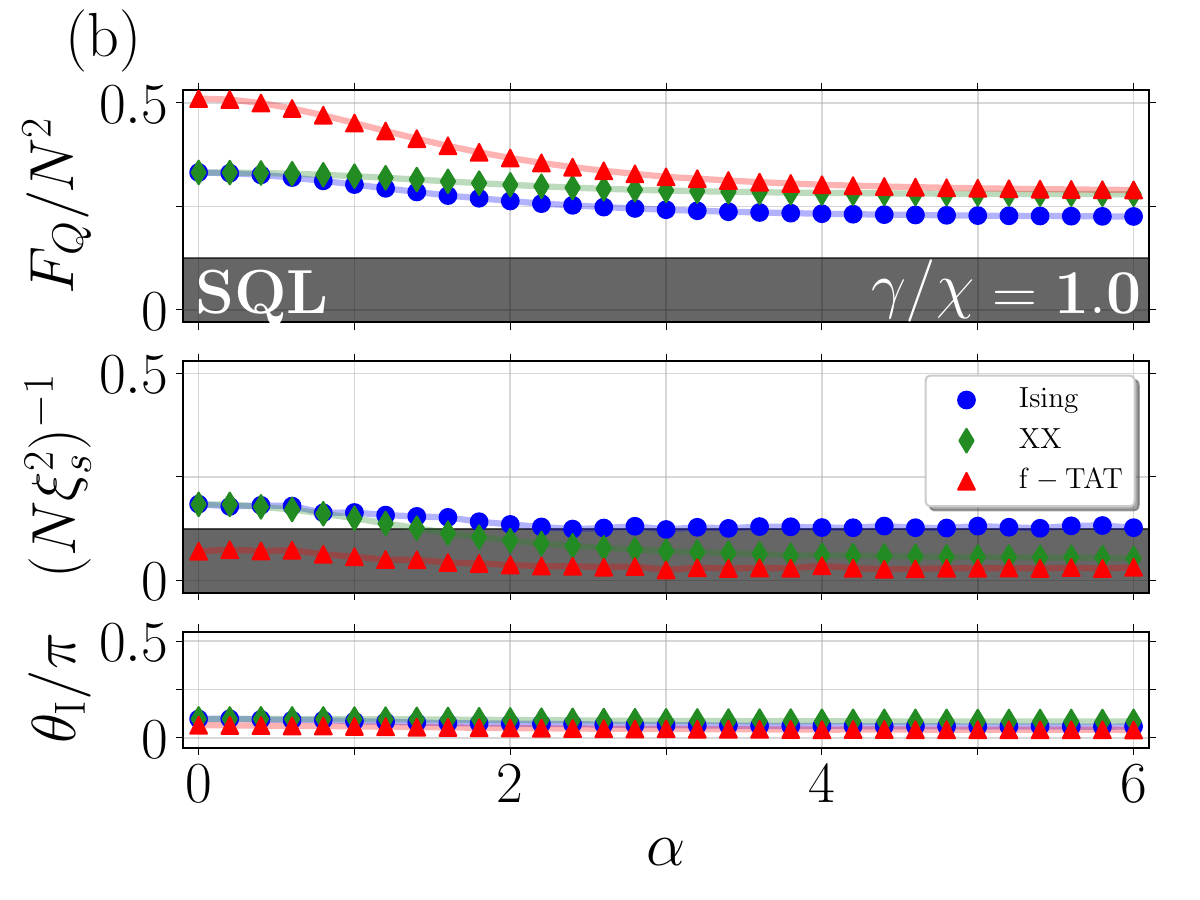}
    \caption{Optimal QFI (top subplots), diagnostic quantities (middle subplots), and interaction parameters $\theta_{\mathrm{I}}$ (lower subplots) as functions of interaction range $\alpha$ for Ising (blue circle), XX (green diamond), and f-TAT models (red triangle). All points are computed with $N = 8$ for choices of (a) $\gamma/\chi= 0.01$ and (b) $\gamma/\chi= 1.0$. Note that we present the QFI and inverse squeezing quantity on a different vertical scale in panel (b).}  
    \label{fig:qfi-v-alpha}
\end{figure} 

We carry out a more systematic investigation of the influence of the interaction range in Fig.~\ref{fig:qfi-v-alpha}, wherein we plot the optimal QFI for each choice of entangling Hamiltonian as a function of $\alpha$. Panels (a) and (b) present results for fixed values of the decoherence rate $\gamma/\chi= 0.01$ and $\gamma/\chi= 1.0$, respectively. 
To further characterize the generated states, we  plot the corresponding VQC parameter $\theta_I$ and (a) the fidelity with a GHZ state ($\mathcal{F}_{\mathrm{GHZ}}$) or (b) the inverse squeezing parameter ($(N\xi^2_s)^{-1}$). All data uses $N = 8$ qubits. 

In the case of weak decoherence [Fig.~\ref{fig:qfi-v-alpha}(a), $\gamma/\chi= 0.01$], the QFI behaves in a markedly distinct manner for each of the models. In the case of the Ising Hamiltonian, the QFI begins near the HL ($F_Q/N^2 \approx 1$) for $\alpha = 0$ and quickly decreases as $\alpha$ increases (i.e., the interaction range shrinks), approaching a value of $F_Q/N^2 \approx 0.27$ (still just above the SQL, $F_Q/N^2 = 1/8$) for $\alpha = 6$. 
The fidelity with a GHZ state (middle panel) similarly drops as the interaction is decreased, which explains the reduction of the QFI relative to the HL. Underpinning both observations, the VQC parameter $\theta_I$ abruptly drops away from values associated with the generation of GHZ-like states ($\theta_I/\pi \sim 0.5$) toward faster evolution that indicates the optimal VQC is preparing squeezed-like states.

% \changed{The abrupt drop in the VQC parameter $\theta_I$ is consistent with the production of squeezed-like states for shorter interaction ranges. The interaction duration deviates from values associated with the generation of GHZ-like states ($\theta_I/\pi \sim 0.5$) and rapidly decreases to values associated with the production of squeezed-like states ($ 0 < \theta_I/\pi < 0.5$). 
% This and a rapidly and monotonically decreasing trajectory for $\mathcal{F}_{\mathrm{GHZ}}$ are consistent with our observations in Fig.~\ref{fig:finite-range-rel-quantities-v-gamma}}, wherein there was an absence of a cat-like regime.}

On the other hand, while for the XX and f-TAT cases the QFI again starts near the HL for $\alpha = 0$, it does not monotonically decrease. Instead, it features a dip for intermediate values of $1 \lesssim \alpha \lesssim 3$ before growing again to become near the HL for the larger values of $\alpha$ we probe (reaching $F_Q/N^2 \approx 0.83$ for $\alpha = 6$). Despite the qualitative similarity, the behavior of $\theta_I$ provides insight that distinguishes the two Hamiltonians. For the XX Hamiltonian, we observe that $\theta_I$ is relatively unchanged as $\alpha$ is varied, suggesting that the existence of the cat-like regime is robust regardless of interaction range \cite{perlin2020spin}. This assertion is supported the fact that the overlap with the GHZ state remains appreciable throughout (i.e., $\mathcal{F}_{\mathrm{GHZ}}$ is above or  close to $1/2$ throughout). By contrast, for the f-TAT Hamiltonian the more substantial dip in the QFI is accompanied by an abrupt drop in the interaction duration $\theta_I$ to smaller values consistent with the generation of squeezed-like states, before jumping back to larger values consistent with the generation of cat-like states. Again, this behavior is supported by a computation of $\mathcal{F}_{\mathrm{GHZ}}$, which similarly falls below $1/2$ in the region of $\alpha$ where smaller $\theta_I$ is obtained before rising back above $1/2$ when $\theta_I$ increases. The behavior of the QFI at large $\alpha$ for both the XX and f-TAT Hamiltonians suggests that the generation of quasi-collective, cat-like states with large QFI is possible with short-range interactions (at least in small systems) and merits further investigation.

The case of more moderate decoherence [Fig.~\ref{fig:qfi-v-alpha}(b), $\gamma/\chi= 1.0$] presents a simpler story, with the QFI starting from smaller values at $\alpha = 0$ and monotonically decreasing as $\alpha$ is increased. In parallel, the values of the associated optimal VQC parameter $\theta_I$  (and the inverse squeezing parameter $(N\xi^2_s)^{-1}$) are consistent with those previously observed in the squeezed-like regime, suggesting it remains robust regardless of $\alpha$.

\subsection{Dependence on qubit number}
Our results so far have focused on the specific case of $N = 8$. However, it is important to examine the dependence of our analysis on system size, particularly to generate insight and develop general strategies for larger systems that are not accessible with exact numerical calculations. 

\begin{figure}
    \centering
    \includegraphics[width=0.8\linewidth]{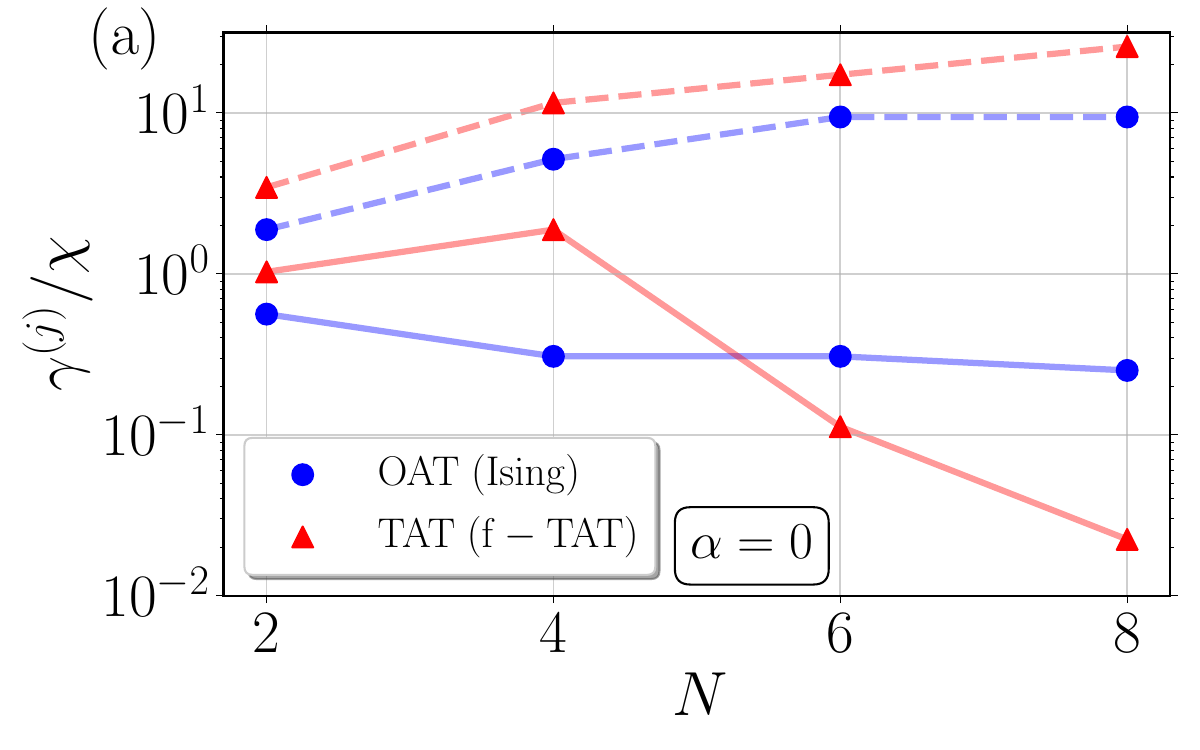}\\
    \includegraphics[width=0.8\linewidth]{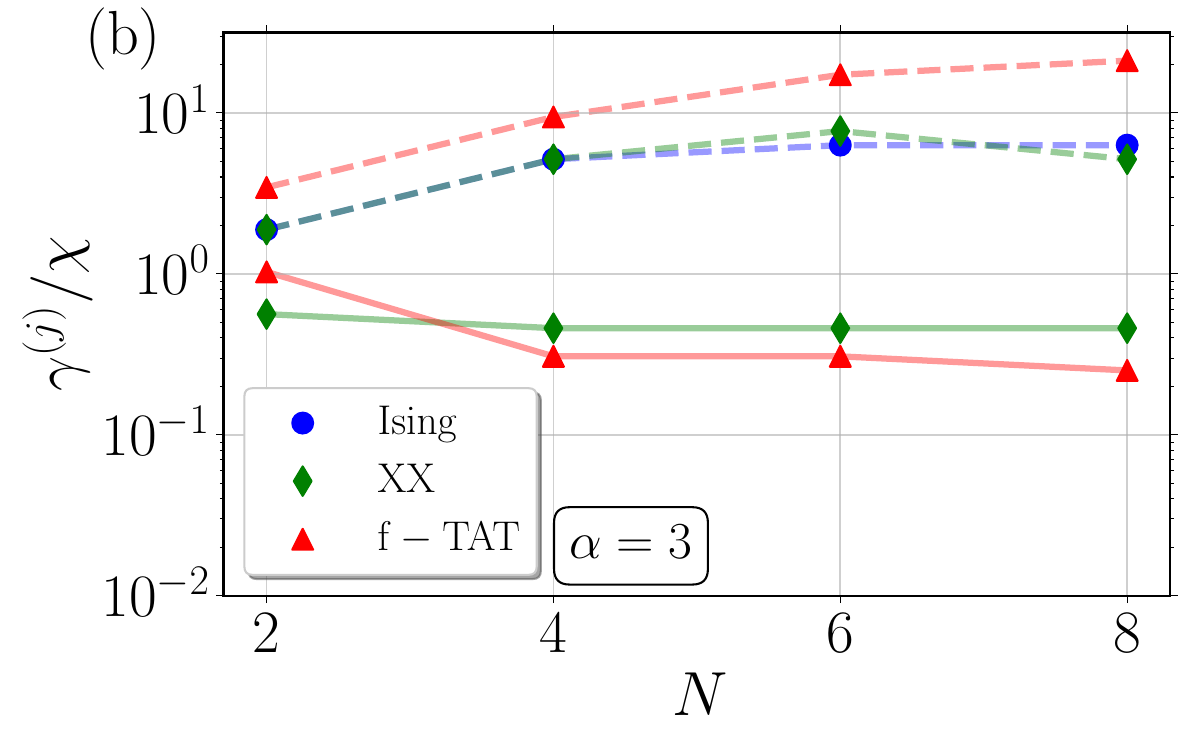}\\
    \includegraphics[width=0.8\linewidth]{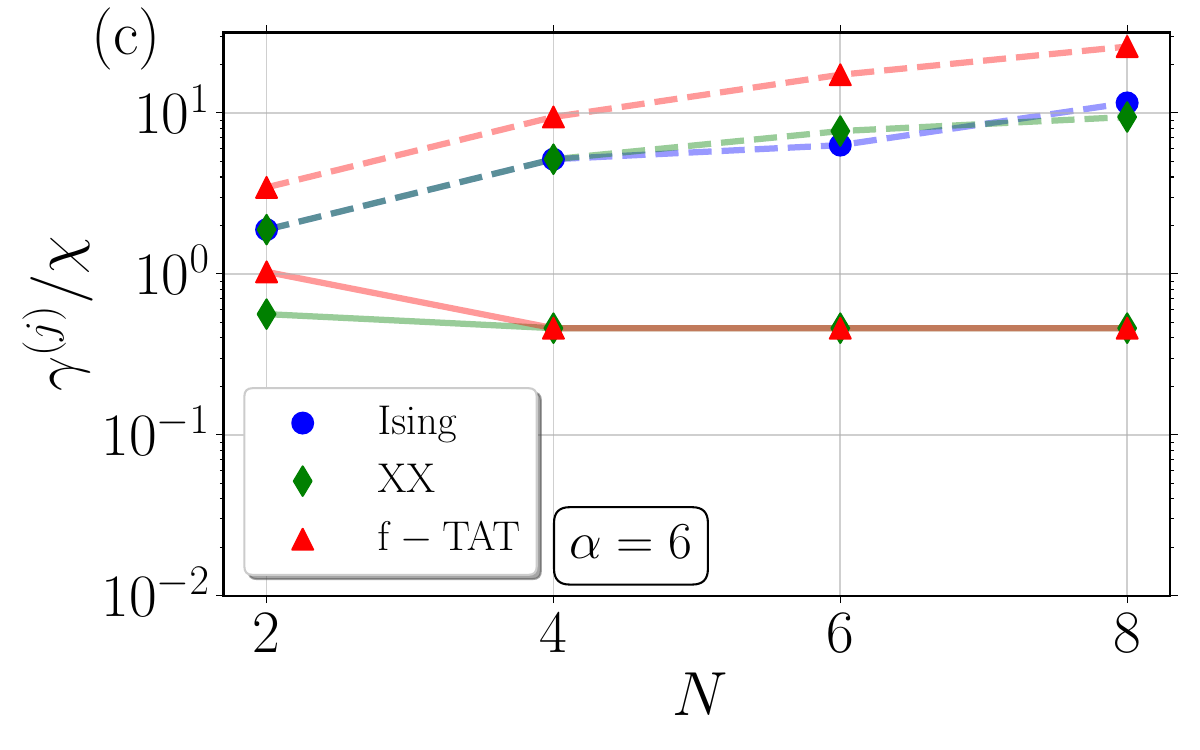}
    \caption{Dephasing strengths $\gamma^{(j)}$ ($j = 1,2$) as a function of $N$ for (a) OAT and TAT with $\alpha = 0$, and Ising, XX, and f-TAT with (b) $\alpha = 3$ and (c) with $\alpha = 6$. Solid and dashed lines depict $\gamma^{(1)}/\chi$ and $\gamma^{(2)}/\chi$, respectively.
    }    
    \label{fig:transitions-v-N}
\end{figure}

We conduct additional calculations for system sizes $N \in \{2,4,6,8\}$ and perform a similar analysis to that described in the preceding sections. In particular, we still observe a general delineation of the optimal preparation strategy into the three regimes: cat-like, squeezed-like,, and uncorrelated. We extract the transition values $\gamma^{(1)}/\chi$ and $\gamma^{(2)}/\chi$ for each system size and plot the results in Fig.~\ref{fig:transitions-v-N} (the solid and dashed lines are used to distinguish the markers for $\gamma^{(1)}/\chi$ and $\gamma^{(2)}/\chi$, respectively). 

Panel (a) of Fig.~\ref{fig:transitions-v-N} shows the results for OAT and TAT. We observe that for both entangling Hamiltonians, the scope of the cat-like regime tends to decrease with system size; in other words, $\gamma^{(1)}/\chi$ trends to smaller values. This is unsurprising given that GHZ-like states are known to be highly fragile and difficult to dynamically prepare in the presence of decoherence, and this fragility worsens with system size \cite{Foss-FeigM2013Dqco-ising-cat-states,Huelga_1997_decoherence}. We note that while $\gamma^{(1)}/\chi$ appears to monotonically decrease with system size for OAT, TAT exhibits more sophisticated behavior. Specifically, TAT depicts a slight increase in $\gamma^{(1)}/\chi$ from $N = 2$ to $4$ before dropping away quickly relative to OAT. We attribute this to the fact that TAT does not prepare explicit GHZ states (in fact the characteristic fidelity $\mathcal{F}_{\mathrm{GHZ}}$ deep in the cat-like regime changes noticeably as $N$ is varied) and we are probing particularly small system sizes. In future work, it would be interesting to exploit the permutational symmetry of the TAT dynamics using efficient numerical methods \cite{Baragiola_2010_permutationalcode} to study the non-unitary dynamics for much larger systems $N\gtrsim 100$, in an effort to better understand the complexity of the cat-like regime. In contrast to the shrinking cat-like regime, the value of $\gamma^{(2)}/\chi$ increases similarly with system size for OAT and TAT. Thus, for large $N$ we expect that, for this VQC, preparing squeezed-like states will become the clear optimal strategy for sensing spin rotations in the presence of modest decoherence.

Panels (b) and (c) of Fig.~\ref{fig:transitions-v-N} depict analogous results for the Hamiltonians in Eqs.~\eqref{eqn:ising-ham}--\eqref{eqn:tat-ham} with a choice of $\alpha = 3$ and $\alpha = 6$, respectively. Intriguingly, in panel (c) for the XX and f-TAT Hamiltonians we observe that $\gamma^{(1)}/\chi$ settles to a fixed value $\gamma^{(1)}/\chi \approx 0.46$ independent of system size for $N \geq 4$ (for the Ising Hamiltonian no cat-like regime exists so we do not plot $\gamma^{(1)}/\chi$). Note that this does not imply that finite-range interactions are more robust to decoherence, since the QFI of the states in this persistent cat-like regime can be much smaller than for infinite-range OAT and TAT (see Fig.~\ref{fig:qfi-params-v-gamma}). On the other hand, the increase of $\gamma^{(2)}/\chi$ with system size for each of the Ising, XX, and f-TAT Hamiltonians closely follows the results for OAT and TAT.

\section{Optimal state preparation with general decoherence}\label{sec:imbalanced-decoherence}
In our investigation so far, we have focused on a relatively specific scenario of isotropic dephasing. Here we broaden the scope of our work and present general results for (i) imbalanced dephasing along each axis and (ii) an alternative scenario where the decoherence is due to a combination of spontaneous emission and dephasing along the $z$-axis. While the presence of the cat-like, squeezed-like, and uncorrelated regimes remains, we find that the quantitative extent of these regimes is sensitively dependent on the specific combination of the entangling Hamiltonian and dominant decoherence mechanism.
%With an understanding for the effect of equivalent $x$-, $y$- and $z$-dephasing, we now turn to examining the general case where dephasing noise is unequal, i.e., $\gamma_x \neq \gamma_y/\chi \neq \gamma_z/\chi$. We also place focus on other physically motivated sources of decoherence. In particular, we survey the effects of spontaneous emission with $z$-dephasing. 
% While the presence of the cat-like, squeezed-like and uncorrelated regimes remains, we find that the characteristic existence of the Hamiltonians [see Eqs.~(\ref{eqn:ising-ham})-(\ref{eqn:tat-ham})] with $\alpha = 0$ and $6$ are more robust to certain noise models than others.

\begin{figure*}
\centering
\includegraphics[width=0.25\linewidth]{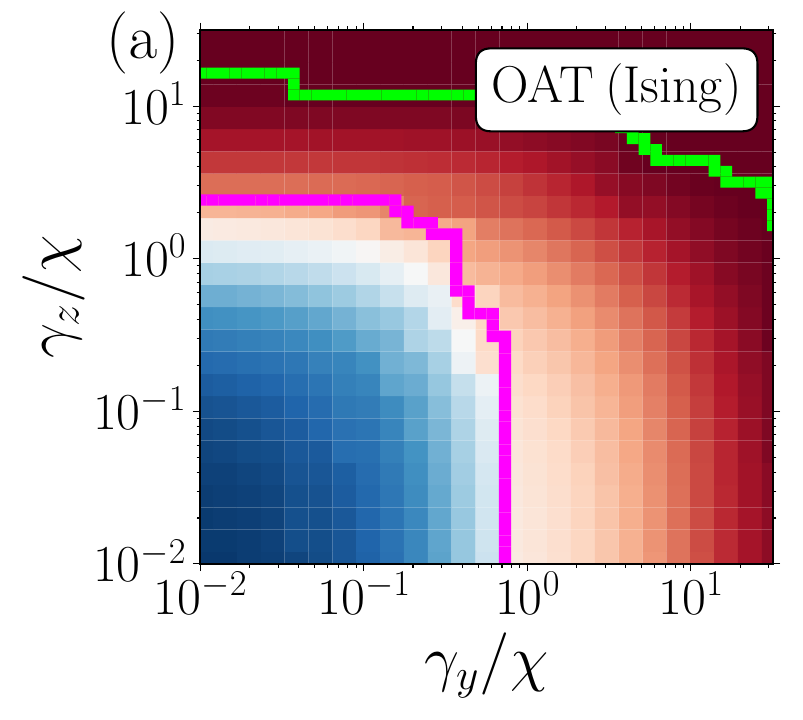}
\includegraphics[width=0.32\linewidth]{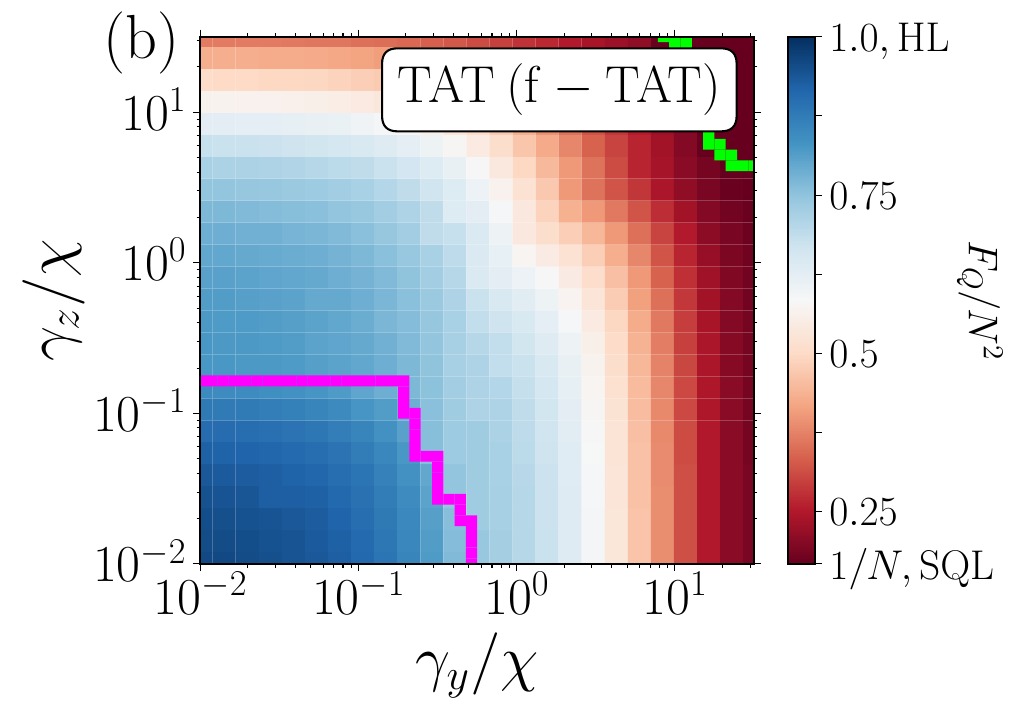}\\
\includegraphics[width=0.25\linewidth]{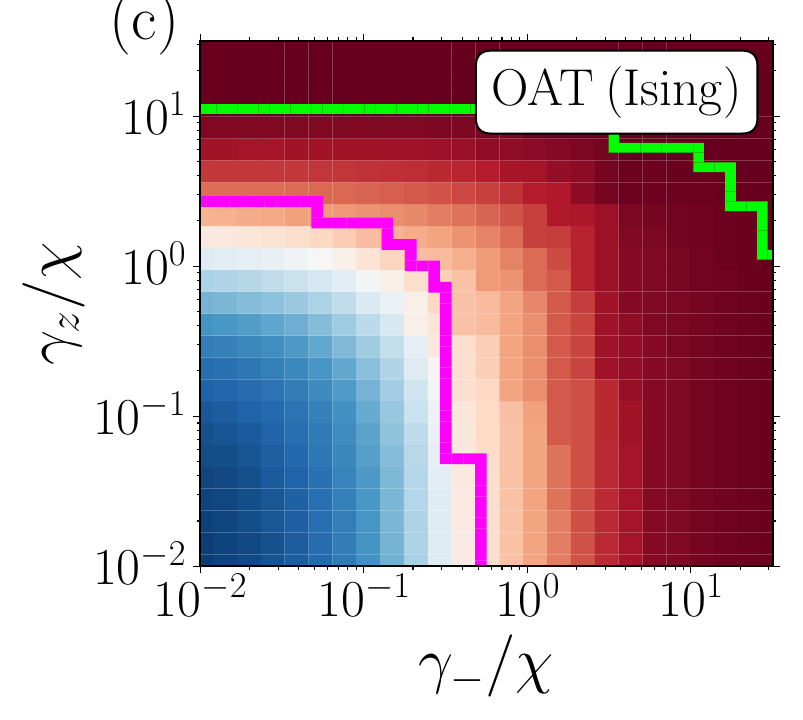}
\includegraphics[width=0.32\linewidth]{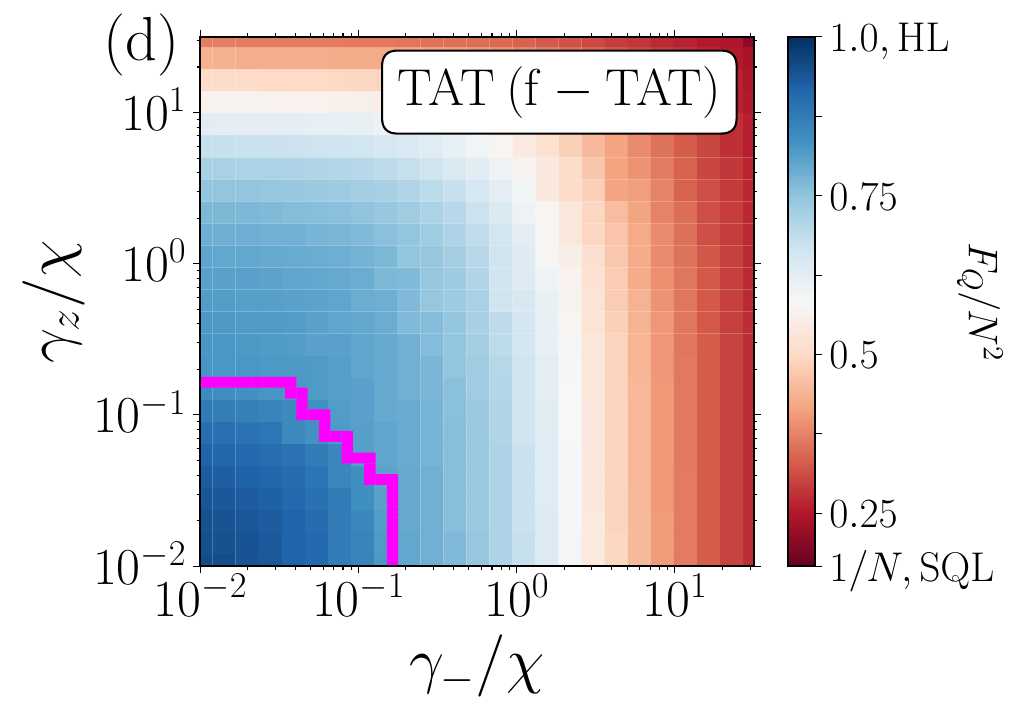}
\caption{Optimal QFI as a function of relative strength of various types of decoherence for OAT in panels (a) and (b) TAT in panels (b) and (d). All data is for $N = 6$ and $\alpha = 0$. The top row scans over $z$-dephasing ($\gamma_z/\chi$) versus $y$-dephasing ($\gamma_y/\chi$) whereas the bottom row scans over $z$-dephasing versus spontaneous emission ($\gamma_-/\chi$). The boundary between the cat-like/squeezed-like regimes and squeezed-like/uncorrelated regimes is delineated by magenta and green lines, respectively.
}
\label{fig:infinite-range-2d-plots}
\end{figure*}

We first present results for the case of $\alpha = 0$ (i.e., OAT and TAT Hamiltonians with infinite-range interactions) with $N=6$ qubits in Fig.~\ref{fig:infinite-range-2d-plots}. Panels (a) and (b) present the optimal QFI ($F_Q/N^2$) for OAT and TAT, respectively, as a function of both $z$-dephasing with strength $\gamma_z/\chi$ and $y$-dephasing with strength $\gamma_y/\chi$ (we ignore $x$-dephasing by setting $\gamma_x = 0$).\footnote{Note that we do not present similar depictions for $x$- versus $z$-dephasing and $x$- versus $y$-dephasing as both $x$- and $y$-dephasing render identical results. A similar statement can be made for the interchanging of spontaneous emission/absorption.} To guide the eye, we indicate the boundaries between the cat-like and squeezed-like regimes with magenta lines and similarly between the squeezed-like and uncorrelated regimes with green lines. The boundaries are defined analogously to $\gammaone$ and $\gammatwo$ indicated previously in the results of Sec.~\ref{sec:results}. 

Starting with OAT in Fig.~\ref{fig:infinite-range-2d-plots}(a), we observe that the extent of the cat-like regime is nearly symmetric, with only a slightly greater sensitivity to $y$-dephasing than $z$-dephasing.
% is relatively symmetric, indicating that it is approximately equally sensitive to both $y$ and $z$-dephasing. 
% \changed{is slightly asymmetric with a greater sensitivity to $y$-dephasing.}
In contrast, we observe that the squeezed-like regime is very asymmetric and is clearly less sensitive to $y$-dephasing. Specifically, for small $\gamma_y/\chi \ll 1$ we find the SQL is reached for $\gamma_z/\chi \sim 10^1$, whereas in the opposing case with small $\gamma_z/\chi \ll 1$ the squeezed-like regime extends far beyond $\gamma_y/\chi \gg 10^1$ [i.e., extends outside the range shown in panel (a)]. In the case of TAT [see \ref{fig:infinite-range-2d-plots}(b)], while we find that the cat-like regime is marginally more sensitive to $z$-dephasing, the more striking observation is the reduced extent of this regime relative to OAT. On the other hand, the squeezed-like regime is much larger than that observed for OAT and notably features larger QFI at comparable values of decoherence. Both observations are consistent with earlier data shown in Fig.~\ref{fig:rel-quantities-v-gamma} for isotropic dephasing.
%In further contrast to OAT, the squeezed-like regime for TAT is also apparently more robust to $z$-dephasing. Specifically, the SQL is not reached until \Com{$\gamma_z/\chi \approx ?$} (not shown) when $\gamma_y/\chi$ is close to zero while appearing much earlier at \Com{$\gamma_y/\chi \approx ?$} (not shown) when $\gamma_z/\chi\approx0$. 

We additionally present results for OAT and TAT subject to spontaneous emission and $z$-dephasing in Fig.~\ref{fig:infinite-range-2d-plots}(c) and (d), where we show the QFI as a function of the respective decoherence rates $\gamma_-/\chi$ and $\gamma_z/\chi$.
% Similarly, results for OAT and TAT as a function of the strength of spontaneous emission and $z$-dephasing are presented in Fig.~\ref{fig:infinite-range-2d-plots}(c) and (d). 
% The cat-like regime for OAT features a \changed{clear} advantage in robustness to $z$-dephasing over spontaneous emission. 
In the case of OAT, while the cat-like regime features a slightly better robustness to $z$-dephasing as opposed to spontaneous emission, the squeezed regime clearly showcases the reverse. The onset of the SQL occurs near $\gamma_z/\chi \approx 10$ when $\gamma_-/\chi$ is taken to be vanishingly small; in contrast,  when $\gamma_z/\chi$ is negligible, the squeezed regime continues to persist for large spontaneous emission $\gamma_-/\chi > 10^{1.5}$ (the crossover the to SQL is not shown). 
% On the other hand, the squeezed-like regime is distinctly more robust to spontaneous emission than $z$-dephasing with the onset of the uncorrelated regime presenting at much earlier values of $z$-dephasing at $(\gamma_-/\chi, \gamma_z/\chi)\approx (10^{-2}, 10^1)$ than spontaneous emission (not shown). 
For TAT [see 7(d)], the extent of the cat-like regime is contracted relative to OAT with the boundary values of $\gamma_-/\chi$ and $\gamma_z/\chi$ (magenta line) effectively reduced by an order of magnitude compared with panel (c). However, the squeezed-like regime again encompasses a considerably larger range of decoherence rates relative to OAT. Specifically, the onset of the uncorrelated regime occurs at large values of $\gamma_z/\chi, \gamma_-/\chi>10^{1.5}$, which are outside the range of parameters displayed in panel (d). In addition, by inspecting the value of the QFI within the squeezed-like regime, we identify  a relative robustness to $z$-dephasing: The onset of red hues (lower QFI) occurs in the limiting case of dominant dephasing at ($\gamma_-/\chi, \gamma_z/\chi) \approx (10^{-2}, 10^{1})$ as opposed $(\gamma_-/\chi, \gamma_z/\chi) \approx (10^0, 10^{-2})$ in the case of dominant spontaneous emission. 

%Idea: Use $(\gamma_y/\chi,\gamma_z/\chi) = ?, i.e., "SQL is reached at the points $(\gamma_y/\chi,\gamma_z/\chi) = ? and $(\gamma_y/\chi,\gamma_z/\chi) = ? on the $\gamma_y/\chi$ and $\gamma_x$ axes, respectively.

% Across panels \ref{fig:infinite-range-2d-plots}(a) and \ref{fig:infinite-range-2d-plots}(b), we observe OAT's cat-like regime is more robust than TAT's for $z$-dephasing while appearing nearly identical for $y$-dephasing. By sharp contrast, the squeezed-like regime for TAT features a broader and more robust area than OAT for both $z$- and $y$-dephasing. Specifically, TAT's and OAT's squeezed-like regimes are more robust for $y$- and $z$-dephasing, respectively. We find qualitatively similar results for $z$-dephasing versus spontaneous emission with strength $\gamma_-/\chi$ across panels \ref{fig:infinite-range-2d-plots}(c) and \ref{fig:infinite-range-2d-plots}(d). Interestingly, OAT features a stronger resilience to $y$-dephasing than spontaneous emission whereas the opposite is true for TAT.

\begin{figure*}
\centering
\includegraphics[width=0.25\linewidth]{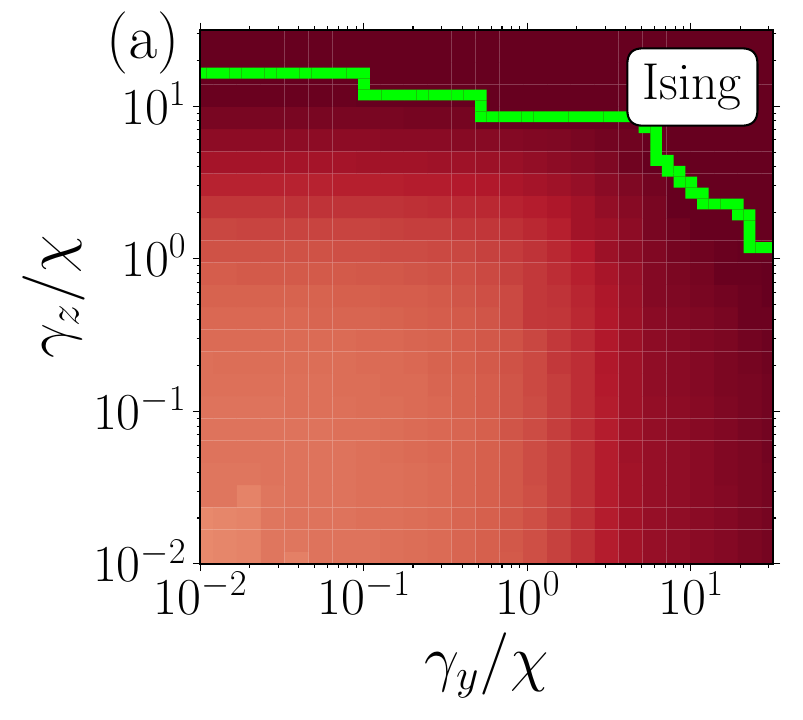}
\includegraphics[width=0.25\linewidth]{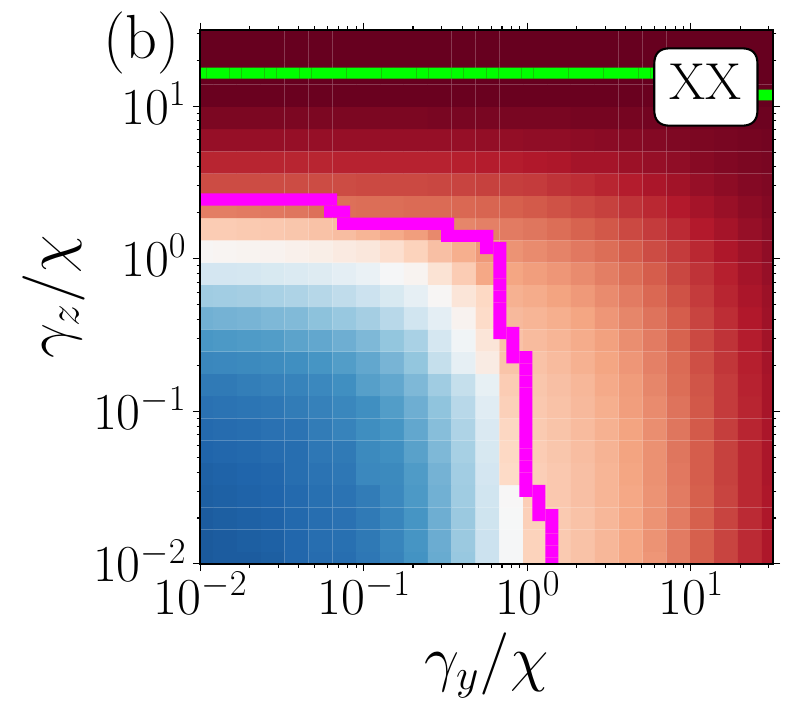}
\includegraphics[width=0.32\linewidth]{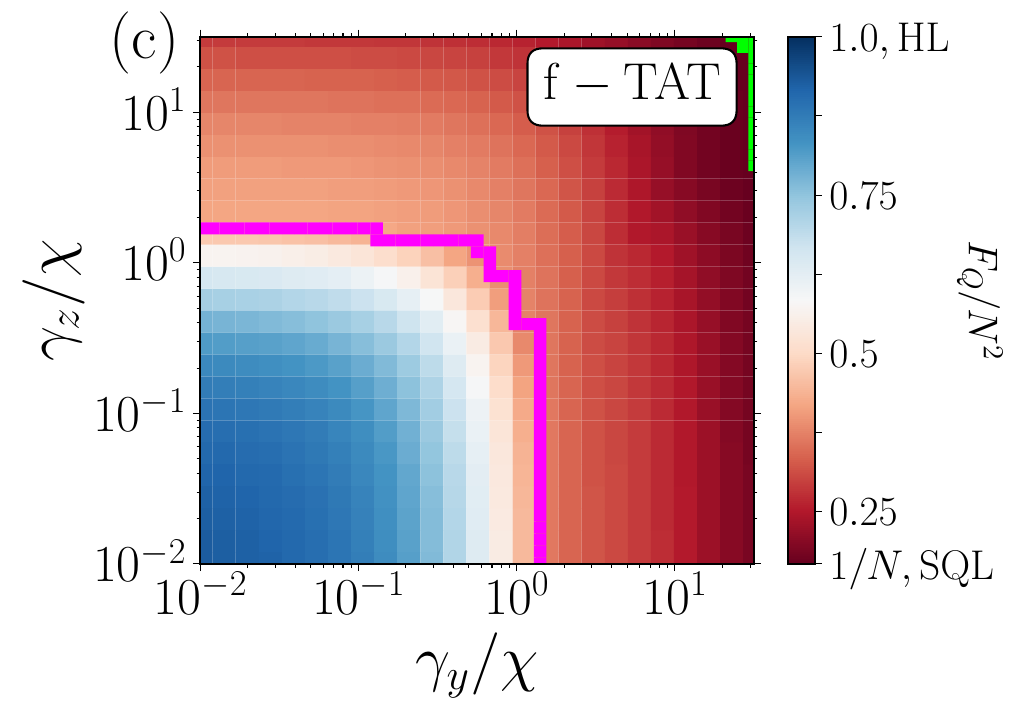}\\
\includegraphics[width=0.25\linewidth]{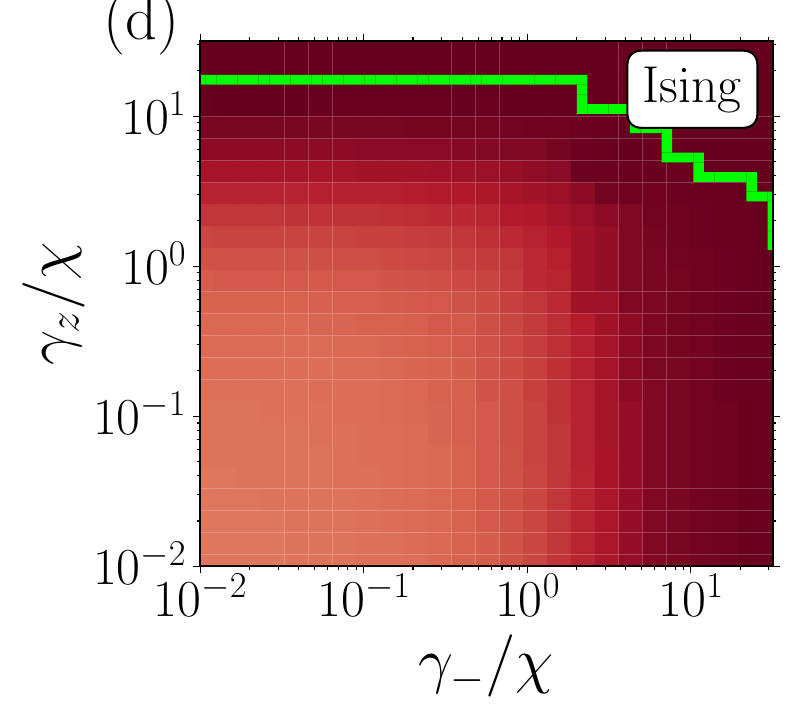}
\includegraphics[width=0.25\linewidth]{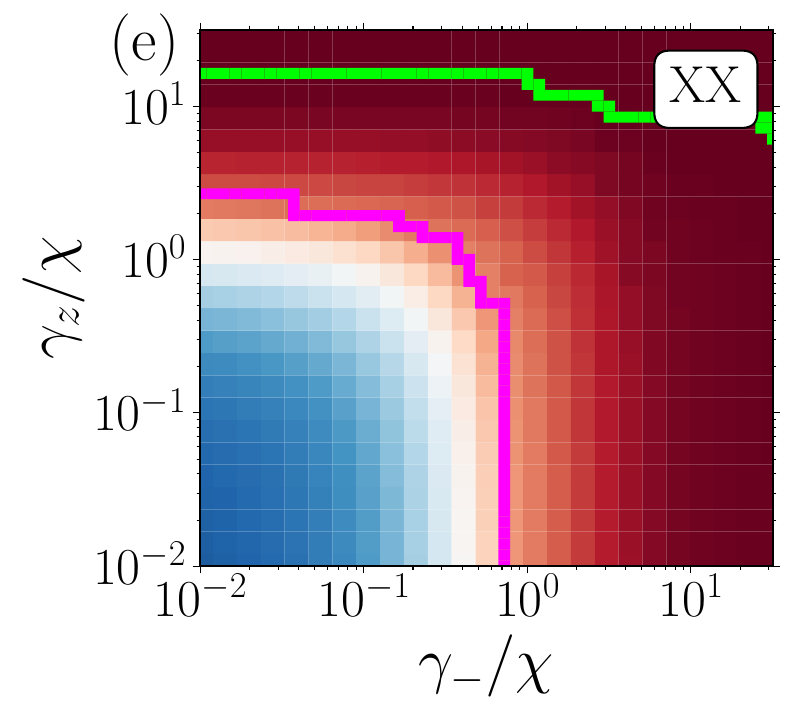}
\includegraphics[width=0.32\linewidth]{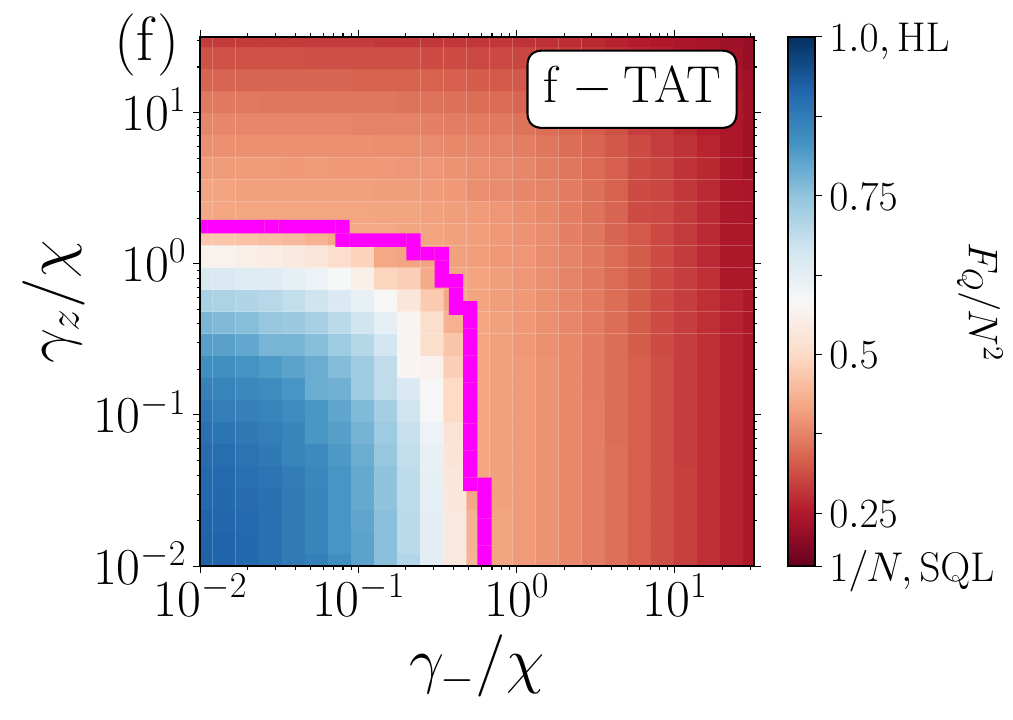}
\caption{Optimal QFI as a function of relative strength of various types of decoherence for Ising [panels (a) and (e)], XX [panels (b) and (e)], and f-TAT [panels (c) and (f)]. All data is for $N = 6$ and $\alpha = 6$. The top row scans over $z$-dephasing ($\gamma_z/\chi$) versus $y$-dephasing ($\gamma_y/\chi$), and the bottom row scans $z$-dephasing versus spontaneous emission ($\gamma_-/\chi$). The boundary between the cat-like/squeezed-like regimes and squeezed-like/uncorrelated regimes are delineated by magenta and green lines, respectively.}
\label{fig:finite-range-2d-plots}
\end{figure*}

% results for the case of $\alpha = 0$ (i.e., OAT and TAT Hamiltonians with infinite-range interactions) with $N=6$ qubits 

Figure \ref{fig:finite-range-2d-plots} presents analogous results for Ising, XX, and f-TAT models with $\alpha = 6$ and $N = 6$ qubits. Panels (a)--(c) display the optimized QFI as a function of dephasing rates $\gamma_y/\chi$, and $\gamma_z/\chi$ ($\gamma_x$ is again set to zero). 
% the QFI ($F_Q/N^2$) as a function of $\gamma_z/\chi$ versus $\gamma_y/\chi$ with Ising, XX, and f-TAT for a choice of $\alpha = 6$ [see panels \ref{fig:finite-range-2d-plots}(a), \ref{fig:finite-range-2d-plots}(b) and \ref{fig:finite-range-2d-plots}(c)]. These plots are similarly generated with fixed $N = 6$ qubits and have similarly defined delineations between the cat-like and squeezed-like regimes (where applicable) and the squeezed-like and uncorrelated regimes. 
For the XX and f-TAT models, the  cat regime is of similar size and approximately symmetric with respect to the dephasing axis. On the other hand, the Ising model does not feature any cat regime, consistent with our prior discussion in Sec.~\ref{sec:results}. 
% The cat-like regime for XX and f-TAT models feature nearly identical areas. Furthermore, consistent with our results in previous sections, the Ising model fails to produce a cat-like regime with short-range interactions. 
Similar to the $\alpha = 0$ results for OAT, the squeezed-like regime of the Ising and XX models extends to large values of $\gamma_y/\chi$ falling outside the plot range and is much more robust to $y$-dephasing. Equivalently, the squeezed-like regime for the f-TAT model qualitatively mirrors the $\alpha = 0$ results for TAT, being far more robust than the Ising and XX models. 
% On the other hand, the squeezed-like regime for f-TAT exhibits greater resilience to $z$-dephasing than the Ising and XX models (both of which exhibit a similar degree of robustness to $z$-dephasing). Conversely, the squeezed-like regime for the XX model is more robust to $y$-dephasing than the f-TAT and Ising models.

Panels (d)--(f) of Fig.~\ref{fig:finite-range-2d-plots} illustrate the behavior of the optimal QFI as a function of $z$-dephasing and spontaneous emission rates $\gamma_z/\chi$ and $\gamma_-/\chi$. Overall, the relative size and existence of the cat-like and squeezed-like regimes follow similar trends to the data of panels (a)--(c). The most significant difference in the case of imbalanced dephasing is the relatively quicker dropoff of the QFI for the XX model when spontaneous emission is dominant, while the converse is true for f-TAT. 

% Alternatively, panels (d), (e) and (f) of Fig.~\ref{fig:finite-range-2d-plots} respectively present results for the Ising, XX and f-TAT models when subjected to $z$-dephasing and spontaneous emission. While nonexistent for Ising, the cat-like regime for the f-TAT and XX models is almost identical. In the squeezed-like regime, f-TAT is clearly the most robust of the three models with markedly stronger resilience to $\gamma_z/\chi$ and $\gamma_-/\chi$. Furthermore, the transition between the squeezed-like and uncorrelated regimes occurs at nearly identical values of $\gamma_z/\chi$ and $\gamma_-/\chi$ for Ising and XX. A broad comparison across OAT- (i.e., Ising and XX) and TAT-based (i.e., f-TAT) models highlights models based on TAT as more robust with larger squeezed-like regimes and with the onset of the uncorrelated regime occurring at larger values of $\gamma_z/\chi$, $\gamma_y/\chi$ and $\gamma_-/\chi$. We attribute this general feature to the exponentially fast dynamics of TAT~\cite{spin-squeezing-ueda} thereby generating states with large QFIs at rates that compete with the decoherence rates for the noise models in question. Nevertheless, OAT tends to feature larger cat-like areas than TAT with a more robust profile along $\gamma_z/\chi$. 

\section{Conclusion} \label{sec:conclusion}
Our work provides insight into optimal state preparation strategies to maximize the QFI for metrology tasks while fully accounting for limitations imposed by decoherence. In particular, we identify a clear delineation of the strategies in terms of the types of states they generate, regardless of the entangling interactions used as part of the VQC. Moreover, while we find that the generation of states with QFI scaling near the HL appears to become increasingly ambitious for larger system sizes, as a result of the increased fragility of the preparation schemes to decoherence, we emphasize that the generation of states with a QFI above the SQL can
% in the presence of decoherence 
scale favorably with $N$. Furthermore, our results suggest that finite-range interactions may be sufficient to generate states with appreciable QFI.

% With a focus on the preparation of states with optimal QFIs within a framework of noisy environments, our results confirm the existence of quantum states with QFIs above the SQL relevant to many fields, e.g., quantum metrology and quantum information science. We inform on the nature of the states generated with a set of physically relevant collective and finite-range Hamiltonians for a given decoherence $\gamma$ and classify them into three regimes closely associated with their QFIs and optimal preparation strategies. While we find that the generation of states with QFIs scaling near the HL appears to be an increasingly ambitious task with increasing system size (on account of rising sensitivities to deleterious noise), we emphasize that the generation of states with QFIs above the SQL
% % in the presence of decoherence 
% scales favorably with $N$. Furthermore, our results indicate promise for the generation of states with QFIs practical toward quantum-enhanced sensing and communication tasks with physical systems featuring limited interaction ranges. 

A number of interesting future directions can build on the analysis presented in this work. First, we can expand the circuit ansatz in Fig.~\ref{fig:VQC} by, for example, increasing the depth of our VQC by layering alternating rotations and entangling stages. This will, in principle, increase the complexity and range of potential states that we can prepare, potentially modifying our current classification into cat-like, squeezed-like, and uncorrelated states.
The VQC ansatz can also be expanded to fully exploit the programmability of quantum sensing hardware by including different site-resolved qubit rotations and two-qubit gates, as opposed to global rotations and entangling operations.

Second, in addition to decoherence introduced during the entangling stage of the VQC, we can incorporate decoherence introduced by errors in VQC parameters such as imperfect initial and final rotations (i.e., shot-to-shot fluctuations in rotation axis or angle). In contrast to the current VQC that we study, this could benefit theoretical understanding by introducing some degree of control over the level of mixedness of the prepared states that is independent of the VQC parameters (whereas in the current protocol, the total amount of decoherence is correlated with the duration of the entangling stage). 

Third, our investigation of the QFI of mixed quantum states can be broadened by considering isolated quantum systems that feature multiple degrees of freedom that can be entangled and coherently controlled but are not necessarily mutually measurable. This leads to a situation where the metrological potential of a prepared quantum state is best captured by the QFI of the reduced density matrix describing the observable degree(s) of freedom, which will be mixed if the system features entanglement \cite{RLS_2024_nonclassicalmotion}. 

We are investigating the development of more sophisticated, specialized numerical optimization routines for optimizing the QFI for a given VQC or state preparation protocol. Whereas general-purpose optimization methods, such as the Nelder--Mead algorithm used to generate the results in this manuscript, are designed to receive only scalar outputs, the QFI form in Eq.~\eqref{eqn:qfi} has considerable structure that can be exploited. Instead of receiving only the scalar value $F_Q[\densitymat ; \hat{G}]$, we are developing optimization methods that can receive and work with the density matrix $\densitymat$ and approximate (finite-differences) or exact (algorithmic-differentiated) derivatives for entries of $\densitymat$ with respect to the parameters $\mathbf{x}$. It is not straightforward to pass this information through an eigendecomposition, which is needed in Eq.~\eqref{eqn:qfi}. However, by building on past work by some of us~\cite{Larson2020,Larson2022} that optimizes composite objectives by using objective knowledge form (such as the form of Eq.~\eqref{eqn:qfi}) in a QFI-only numerical optimization routine, we hope to find higher-quality solutions in fewer queries/simulations of the system. This is a particularly important task because simulation of open quantum systems involving many qubits and subsequent computation of the QFI for a given mixed quantum state are computationally expensive. 

\section*{Acknowledgments}
This material is based upon work supported by the U.S. Department of Energy, Office of Science, Advanced Scientific Computing Research, Exploratory Research for Extreme-Scale Science program. The majority of computing for this project was performed at the OU Supercomputing Center for Education \& Research (OSCER) at the University of Oklahoma (OU).

%\bibliography{library}

%merlin.mbs apsrev4-1.bst 2010-07-25 4.21a (PWD, AO, DPC) hacked
%Control: key (0)
%Control: author (8) initials jnrlst
%Control: editor formatted (1) identically to author
%Control: production of article title (-1) disabled
%Control: page (0) single
%Control: year (1) truncated
%Control: production of eprint (0) enabled
 \newcommand{\noop}[1]{}

\appendix

\section{Husimi probability distributions}
\label{sec:husimi}

Here we define the Husimi probability distributions used to visualize spin states in Fig.~\ref{fig:rel-quantities-v-gamma}.
The basic idea is to define a Husimi probability distribution within each manifold of fixed spin length $S$ and take an appropriately weighted average of these distributions.
Note that, for ease of language, in this appendix we treat the symbol $S$ as a variable to denote spin length, whereas the main text defines $S=N/2$ for $N$ spins.

\subsection{A simplified case}

Consider first the simplified case in which $N$ spins occupy the permutationally symmetric manifold with spin length $S=N/2$.
In this case, for a density matrix $\hat\rho$ we can define the Husimi probability distribution
\begin{align}
  \rho(\bm v) = w_S \braket{\bm v|\hat\rho|\bm v}
\end{align}
\begin{align}
  \ket{\bm v} = \hat R(\bm v) \ket{\uparrow}^{\otimes N},
  &&
  \hat R(\bm v) = e^{-i\phi\hat S_z} e^{-i\theta\hat S_y},
\end{align}
and the normalization factor $w_S$ is determined by requiring that $\rho(\bm v)$ be a normalized probability distribution, with
\begin{align}
  \int \text{d}\bm v\, \rho(\bm v)
  = \int_0^\pi \text{d}\theta\, \sin\theta
  \int_0^{2\pi} \text{d}\phi \, \rho(\theta,\phi)
  = 1,
\end{align}
which implies that\footnote{The normalization factor $w_S$ is most easily determined by considering a uniform mixture of all $N+1$ permutationally symmetric states, for which $\braket{\bm v|\hat\rho|\bm v}=1/(N+1)=1/(2S+1)$.}
\begin{align}
  w_S = \frac{2S+1}{4\pi}.
\end{align}
When the state $\hat\rho$ is permutationally symmetric, the probability distribution $\rho$ is a faithful representation of $\hat\rho$ in the sense that $\hat\rho$ is uniquely determined by $\rho$.
Moreover, in this case the values $\rho(\bm v)$ at $(2S+1)^2$ points $\bm v$ suffice to reconstruct $\hat\rho$ \cite{perlin2021spin}.

\subsection{The general case}

More generally, the state $\hat\rho$ may have components with non-maximal spin length $S$.
In this case, the state $\hat\rho$ can no longer be faithfully represented by a probability distribution on a sphere.
Nonetheless, we meaningfully visualize the spin polarization of $\hat\rho$ by averaging over probability distributions that represent components of $\hat\rho$ within fixed-$S$ manifolds.
To this end, we classify states by their spin length $S$, spin projection $m$ onto a quantization axis, and an auxiliary index $\xi$ that encodes how a state transforms under spin permutations.
That is, we identify states $\ket{S,m,\xi}$ that are simultaneous eigenstates of $\hat{S}^2 = \hat S_x^2 + \hat S_y^2 + \hat S_z^2$ and $\hat S_z$, with $\hat{S}^2\ket{S,m,\xi}=S(S+1)\ket{S,m,\xi}$ and $\hat{S}_z\ket{S,m,\xi}=m\ket{S,m,\xi}$.
We then define the rotated state
\begin{align}
  \ket{\bm v,S,\xi} = \hat R(\bm v) \ket{S,S,\xi},
\end{align}
which is the analogue of $\ket{\bm v}$ in a sector of Hilbert space with spin length $S$.
The net spin-polarization probability distribution $\rho(\bm v)$ is then
\begin{align}
  \rho(\bm v) = \sum_{S,\xi} \, w_S \braket{\bm v,S,\xi|\hat\rho|\bm v,S,\xi}.
\end{align}

\end{document}